# Long-term option pricing with a lower reflecting barrier

R. Guy Thomas

School of Mathematics, Statistics and Actuarial Science, University of Kent, Canterbury CT2 7FS, UK

E-mail: r.g.thomas@kent.ac.uk

## Abstract

This paper considers the pricing of long-term options on assets such as housing, where either government intervention or the economic nature of the asset is assumed to limit large falls in prices. The observed asset price is modelled by a geometric Brownian motion ("the notional price") reflected at a lower barrier. The resulting observed price has standard dynamics but with localised intervention at the barrier, which allows arbitrage with interim losses; this is funded by the government's unlimited powers of intervention, and its exploitation is subject to credit constraints. Despite the lack of an equivalent martingale measure for the observed price, options on this price can be expressed as compound options on the arbitrage-free notional price, to which standard risk-neutral arguments can be applied. Because option deltas tend to zero when the observed price approaches the barrier, hedging with the observed price gives the same results as hedging with the notional price, and so exactly replicates option payoffs. Hedging schemes are not unique, with the cheapest scheme for any derivative being the one which best exploits the interventions at the barrier. The price of a put is clear: direct replication has a lower initial cost than synthetic replication, and the replication portfolio always has positive value. The price of a call is ambiguous: synthetic replication has a lower initial cost than direct replication, but the replication portfolio may give interim losses, and so the preferred replication strategy (and hence price) of a call may depend on what margin payments need to be made on these losses.

**Keywords:** Long-term option pricing; No-negative-equity guarantee; Put option; Arbitrage; Reflecting barrier

## 1. Introduction

Thomas (2021) considered the valuation of no-negative equity guarantees (NNEG), essentially a form of long-term put option, in the presence of a lower reflecting barrier under the spot price. The pricing method was to integrate the option's intrinsic value over a drift-neutralised density for the asset price at maturity. In Appendix B of that paper, I noted a puzzle: when the same method is applied to price a call option, the resulting pairs of call and put prices for the same strike do not satisfy the standard put-call parity. I suggested a number of pragmatic reasons why we should nevertheless adopt the price of the put for NNEG valuation, and assign some residual ambiguity to the price of the call. Similarly, in an antecedent paper, Hertrich (2015, p239-240) writes that in the presence of a lower reflecting barrier *"…the put price is unique and equals the risk-neutral price…the call market price can take on at least these two values, i.e., the risk-neutral call price and the imputed call price via the standard put-call parity."*

The failure of standard put-call parity hints that the barrier model allows some sort of arbitrage. Previous papers (Thomas (2021), Hertrich and Zimmermann 2017, Hertrich 2015, Veestraeten 2008, 2013) all noted that the assumed instantaneous nature of reflection rules out the obvious arbitrage of buying when the spot price is exactly at the barrier, and then selling a moment later after it rises. However, this argument renders impractical only immediate arbitrage (arbitrage with no interim losses); it does not exclude arbitrage with interim losses, which requires credit to cover losses which may be incurred before an eventual gain.

Strategies of this type are indeed feasible in the barrier model. This raises economic and mathematical issues, which were not addressed in any of the previous papers just cited. Economically, it implies that the realism of the assumed price process depends on the government's unlimited powers of intervention, and/or credit limits or other institutional features that limit scaling of the arbitrage. Mathematically, the existence of any arbitrage strategy, irrespective of its practicality of implementation, implies that no equivalent martingale measure for the asset exists, and therefore that the usual risk-neutral pricing arguments cannot be applied, at least in their standard form.

Nevertheless, delta hedging using the formulas in previous papers does exactly replicate option payoffs. The reasons are as follows. Although geometric Brownian motion (GBM) reflected at a lower barrier ("the observed price") is not arbitrage-free, the *underlying* geometric Brownian motion ("the notional price") is arbitrage-free. Furthermore, options on the observed price can be expressed as compound options on the notional price, and then expectations taken under a risk-neutral measure for the notional price. Although the notional price is not available for hedging, it turns out that for the hedging schemes we need to operate, hedging with the observed price achieves the same results as hedging with the notional price. This is because the deltas for options always tend to zero as the observed price approaches the localised irregularity at the barrier; and everywhere else, the observed price has the same standard GBM dynamics as the notional price.

The resulting replicating portfolios are martingales under the risk-neutral measure for the notional price. But the observed price is a sub-martingale under this measure, because of the interventions at the barrier (which push the observed price upwards, and are not cancelled by the removal of the drift in the standard risk-neutralisation). Similarly, the cheapest replication of a forward contract is a sub-martingale; this is because the cheapest replication of a forward contract has a static positive delta, and so is positively exposed to the interventions at the barrier. These different dynamics of replicating portfolios – options martingale, forward contracts sub-martingale – inevitably mean that call, put and forward prices together will not satisfy the standard put-call parity.

Put-call parity can be restored by invoking the construct of synthetic replication for one of the options, in contrast with the more common direct replication. For a call, direct replication involves a continuously varying portfolio which is long a fraction of the underlying asset, and short cash. Synthetic replication involves two elements: a long position in a forward contract, and a long position in a put, both with the same strike as the call we wish to replicate. The synthetic call price is then determined as the lowest cost of replicating each of these two elements separately (via a static position for the forward, and a dynamic position for the put). An analogous synthetic replication can be constructed for a put.

Under standard models for the asset price, direct replication and synthetic replication have the same costs. But in the presence of the reflecting barrier, their costs can differ. I argue that arbitrage will tend to enforce the lowest replication cost as the price for any derivative (I call this "the principle of the lowest price", in contrast to the usual "law of one price"). For a put, direct replication always has a lower initial cost, and never produces interim losses, so direct replication is unambiguously the best way of replicating a put. For a call, synthetic replication always has a lower initial cost, but it produces interim losses, and so the preferred replication strategy (and hence price) of a call may depend on what margin payments need to be made on these losses.

If the price of a call is ambiguous, then put-call parity is also in general ambiguous. But we can define it if we assume that no margin payments are required on interim losses. In that case, synthetic replication becomes unambiguously preferable for a call, so put-call parity then takes the form *Synthetic Call price – Put price = Forward Contract price.*

Earlier work on option pricing in the presence of a lower reflecting barrier includes the following papers. For calls with a lower barrier, Veestraeten (2008) derived the direct-replication price, and Veestraeten (2013) considered a combination of upper and lower barriers to represent a foreign exchange currency band. For puts with a lower barrier, Hertrich and Veestraeten (2013) stated the result, and Hertrich (2015) and Hertrich and Zimmermann (2017) gave a full derivation. Thomas (2021) suggested an application to the valuation of no-negative-equity-guarantees on equity release mortgages. Each of these papers focused mainly either on call options (Veestraeten 2008, 2013) or on put options (Hertrich 2015, Hertrich and Zimmermann 2017; Thomas 2021), but not both; this may explain why the complications highlighted in the present paper have not previously been considered. In earlier actuarial literature, Gerber and Pafumi (2000) and Imai and Boyle (2001) considered the related problem of pricing of a dynamic guarantee on an investment fund, where the guarantee is modelled as a reflecting barrier under the fund price. Ko et al. (2010) considered an investment fund with two reflecting barriers, representing a lower guarantee plus withdrawals to prevent the fund from exceeding an upper limit. The formulas for these related problems can be reconciled with those in this paper, see e.g. Appendix D of Thomas (2021). Buckner et al. (2022) draw attention to the arbitrage opportunities created by a reflecting barrier; I agree with this observation, but not with the inference that the option formulas have no justification.

The conclusions of this paper can be compared with those in the literature on option prices in the presence of a bubble in the price of the underlying asset (e.g. Cox and Hobson 2005, Heston et al. 2007, Jarrow et al. 2007, Protter 2013). In both settings, the standard put-call parity fails to hold. In a bubble, the current asset price exceeds its discounted expectation under the pricing measure at a future date; but with a lower reflecting barrier, this inequality is reversed. Protter (2013) distinguishes between "fundamental" and "market" prices for options in the presence of a bubble, where the fundamental price is the discounted expectation under the risk-neutral measure. For a put, market and fundamental prices in a bubble are the same (as in the barrier model); but for a call, market and fundamental prices differ by an amount equivalent to the difference *Synthetic Call – Call* in this paper (Protter 2013, Equation 80). His difference has the opposite sign to ours: in a bubble, the current market price of the asset (or a call on the asset) exceeds its discounted expectation under the pricing measure, but with a lower reflecting barrier, the inequality is reversed.

The rest of this paper is organised as follows. Section 2 gives the set-up of the barrier model and the derivation of option prices and deltas, and reports simulations showing satisfactory convergence of pricing errors and replication errors. Section 3 considers put-call parity, and hence direct and synthetic replication, and concludes that a put should always be priced by direct replication, but the price of a call is ambiguous. Application of the points made in Section 3 is illustrated in Section 4 for a put, and Section 5 for a call. Section 6 considers the arbitrage possibilities in the barrier model, and their susceptibility to credit limits and other institutional constraints. Section 7 gives further discussion, and Section 8 states conclusions.

**2. The barrier model**

2.1 Concept

The barrier is a metaphor for policymaker actions, such as interventions to support prices after a large fall in the housing market (Thomas 2021, section 2), or central bank currency purchases supporting a floor under an exchange rate (Hertrich, 2015, Hertrich and Zimmermann, 2017). Alternatively, the barrier represents the different economic properties of underlying assets such as freehold land (an absolute claim) compared with the underlying assets such equity (a residual claim) as in many other option valuations (Thomas 2021, section 3).

We start with a standard geometric Brownian motion for the price process. We then impose a reflecting barrier somewhere below the lower of the spot price and the strike price (i.e. $b < \min(S, K)$). The strike can be either lower or higher than the spot price, i.e. in or out of the money. If the spot price hits the barrier, reflection occurs instantaneously, and with infinitesimal size. We can think of this as the State making a small purchase, just sufficient to prevent the price falling below the barrier. The instantaneous nature of the reflection means that the price does not spend any finite time at the barrier, so the obvious arbitrage of buying when the price is exactly at the barrier is not practically feasible (but there will be more to say about arbitrage later).

2.2 Derivations

The derivations make use of two price processes: a process for the observed price $S_t$ in a world with a reflecting barrier; and a process for the notional price $N_t$ of an asset with the same starting value today and same instantaneous dynamics, but in a world without a barrier.

We assume that the notional price $N_t$ is a geometric Brownian motion, specified as usual by

$$dN_t = (\mu - q) N_t \, dt + \sigma N_t \, dW_t \tag{1}$$

where $\mu$ is the real-world expected total return on the asset, $q$ is the yield (assumed to be continuous), $\sigma$ is the diffusion and $dW_t$ is the increment of a standard Wiener process. The solution of this is

$$N_t = N_0 \exp\{(\mu - q - \sigma^2/2)t + \sigma W_t\} \tag{2}$$

where $N_0$ is the notional price today.

Now consider a lower reflecting barrier at price level $b$. Because of the barrier, the observed price process $S_t$ will be

$$S_t = N_t \cdot \max\left[1, \max_{0 \leq n \leq t}\left(\frac{b}{N_n}\right)\right]. \tag{3}$$

In words: the process $S_t$ is equal to $N_t$, if $N_t$ has never gone below the barrier; or an up-rated version of $N_t$ which "undoes" the maximum proportional deficit relative to the barrier, if it has. Figure 1 shows how it works, for one unfavourable simulation over 25 years.

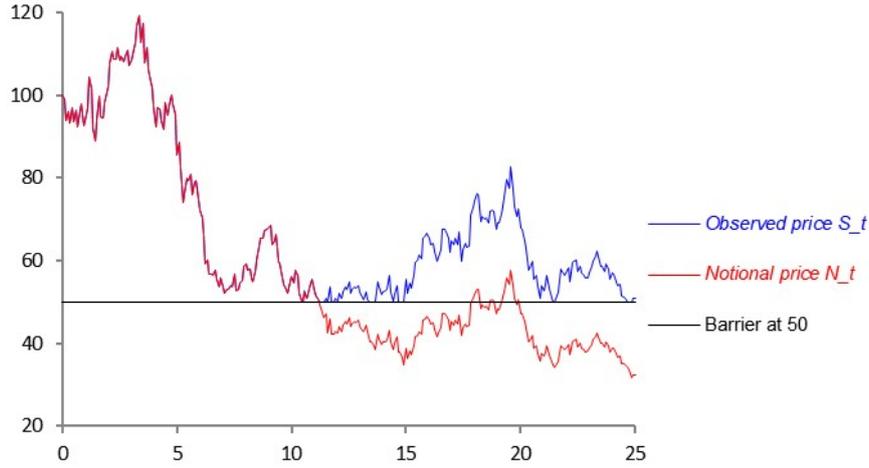

**Figure 1.** One simulation for the prices $N_t$ and $S_t$

Now transform to an arithmetic Brownian motion: divide Equation 3 by $b$ (so that the barrier becomes 1), and then take logs (so that the barrier becomes zero).[1] This gives:

$$\ln\left(\frac{S_t}{b}\right) = \ln\left(\frac{N_t}{b}\right) + \ln\left(\max\left[1, \max_{0 \leq n \leq t}\left(\frac{b}{N_n}\right)\right]\right) \quad (4)$$

or, writing $Z_t = \ln(S_t/b)$, $Y_t = \ln(N_t/b)$, and noting $\ln(\max(a,b)) = \max(\ln a, \ln b)$, and $N_n = b \exp Y_n$:

$$Z_t = Y_t + \max\left[0, \max_{0 \leq n \leq t}(-Y_n)\right] \quad (5)$$

or

$$Z_t = Y_t + L_t. \quad (6)$$

Figure 2, which is the same simulation as in Figure 1 but with the transformation just given, shows how it works. Note that the blue line $Z_t$ is obtained by taking the red line $Y_t$ and then adding $L_t$, the running maximum shortfall of $Y_t$ below zero. $L_t$ is a non-decreasing continuous process, which increases whenever $Z_t$ touches the barrier. This construction ensures that $Z_t$ never spends any finite time at the barrier, and there are no jumps. $L_t$ can also be thought of as the total quantum of the

---

[1] Strictly, we only need to take logs to give an arithmetic Brownian motion. The prior division by $b$ just translates the barrier for the logged variable to zero, and so gives a more intuitive appearance in Figure 2.

interventions at the barrier between time 0 and time *t*; this quantum is finite, despite the individual interventions being infinitesimal.[2]

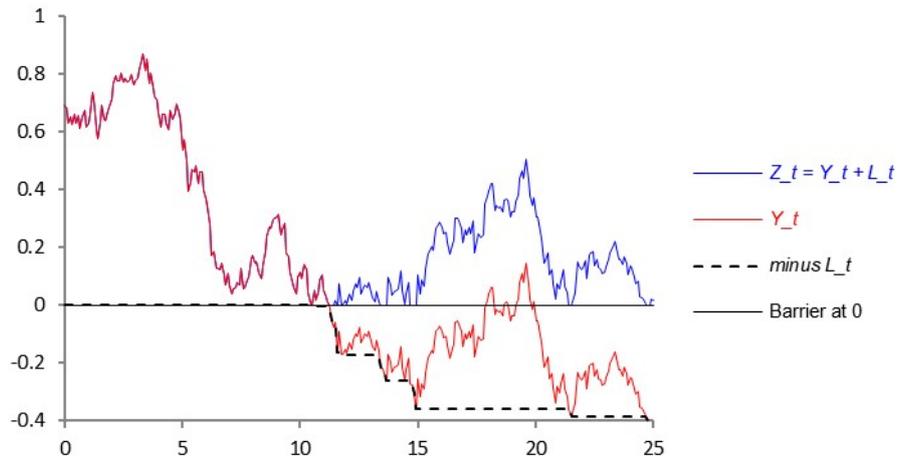

**Figure 2.** One simulation for the log prices $Y_t$ and $Z_t$

Now note that $Y_t$ is an arithmetic Brownian motion, for which Ito's lemma gives

$$dY_t = d\ln(N_t/b) = d\ln N_t = (\mu - q - \sigma^2/2)dt + \sigma dW_t \qquad (7)$$

Then note that $Z_t$ is a semi-martingale, because it can be decomposed into a martingale (the $W_t$ term) and two finite variation processes, the drift and the reflection. So we can also apply Ito's lemma to $Z_t = f(t, W, L)$. This gives

$$dZ_t = d\ln(S_t/b) = d\ln S_t = (\mu - q - \sigma^2/2)dt + \sigma dW_t + dL_t \qquad (8)$$

which is the same as Equation 7, plus $dL_t$, and second order terms involving $dLdW$, $dLdt$, and $dLdL$ which are all zero (because of the infinitesimal size of the increments at reflection).

Comparing Equations 7 and 8, we see that reflected Brownian motion $Z_t$ has the same stochastic process as the ordinary Brownian motion $Y_t$, except for the extra $dL_t$ term. This term is zero at all times, except when $Z_t$ touches the barrier (equivalently: $Y_t$ makes a new low below zero), where it generates an infinitesimal increment.

---

[2] The notation $L_t$ arises because of a technical property of the stochastic process: the running maximum shortfall of $Y_t$ below zero is equivalent to 0.5 times the "local time" of $Z_t$ at zero, where local time is a measure of the total amount of time between 0 and *t* that $Z$ spends "very close" to zero.

The equivalent equations for the non-logged variables are

$$dN_t = (\mu - q)N_t\, dt + \sigma N_t\, dW_t \qquad (9)$$

and

$$dS_t = (\mu - q)S_t\, dt + \sigma S_t\, dW_t + dL_t. \qquad (10)$$

By replacing the expected total return $\mu$ in Equation 9 with the risk-free rate $r$, we get a process for $N_t$ for a new measure $\mathbb{Q}_N$, under which the geometric Brownian motion $N_t$ discounted at $r - q$ is a martingale:

$$dN_t = (r - q)N_t\, dt + \sigma N_t\, dW_t^{\mathbb{Q}_N} \qquad (11)$$

where $W_t^{\mathbb{Q}_N}$ is a Wiener process under the new measure $\mathbb{Q}_N$.[3]

Earlier papers (Thomas 2021, Hertrich and Zimmermann 2017, Hertrich 2015, Veestraeten 2008, 2013) implicitly proceeded on the basis that this change of measure also turned the discounted observed price $S_t$ into a martingale. But this is not correct, because of the $dL_t$ term in Equation 10. The discounted notional price (red line in Figure 1) is a martingale under $\mathbb{Q}_N$ (i.e. $e^{-(r-q)T}\mathrm{E}_{Q_N}[N_T] = S_0$); but the discounted observed price (blue line) tends to be pushed upwards at the barrier, and so is a sub-martingale under $\mathbb{Q}_N$ (i.e. $e^{-(r-q)T}\mathrm{E}_{Q_N}[S_T] \geq S_0$).

The earlier papers obtained pricing formulas for options on $S_t$ as discounted $\mathbb{Q}_N$-expectations of terminal payoffs. The usual justification is that because the Black-Scholes partial differential equation for the price of an option does not depend directly on investor preferences, its solution must be consistent with the expected value of the option in a world of risk-neutral preferences. In the Black-Scholes model, we represent risk-neutral preferences by setting the expected total return on the asset to be the same as the risk-free rate $r$; or stated differently, changing the measure to make the stochastic process for the discounted asset price a martingale. But in the presence of the barrier, we have already noted that discounted $S_t$ is *not* a martingale under $\mathbb{Q}_N$ (see previous paragraph). Furthermore, because of the interventions at the barrier, there is no change of drift that can make it so, and hence the usual procedure does not work. Nevertheless, it turns out that $\mathbb{Q}_N$-expectations do give hedging schemes which exactly replicate option payoffs, albeit for subtly different reasons from the usual argument. These reasons will be explained in Section 2.4 below; but first, I state the $\mathbb{Q}_N$-expectations and the corresponding deltas as in Thomas (2021).

---

[3] The notation $\mathbb{Q}_N$ (instead of the standard $\mathbb{Q}$) emphasises that this is a risk-neutral measure for the notional price $N_t$, not the observed price $S_t$.

For a call, the $\mathbb{Q}_N$-expectation gives a price of

$$C_B = e^{-rT} E_{\mathbb{Q}_N}[S_T - K]^+ = e^{-rT} \int_K^\infty (S_T - K) f(S_T) \, dS_T \qquad (12)$$

where the $B$-subscript on $C_B$ denotes that this is a call in the presence of a lower reflecting barrier. Conveniently, an expression is available for the density of $Z_t$ as a function of $Y_t$, and hence also the re-transformed density for $S_t = b \exp Z_t$, the observed price.[4] This enables us to evaluate the integral as follows (for a full derivation, see Veestraeten, 2008):

$$C_B = S e^{-qT} \Phi(z_1) - K e^{-rT} \Phi(z_1 - \sigma\sqrt{T}) + \frac{1}{\theta} \left\{ \begin{array}{l} +S e^{-qT} \left(\dfrac{b}{S}\right)^{1+\theta} \Phi(z_2) \\ -K e^{-rT} \left(\dfrac{b}{K}\right)^{1-\theta} \Phi(z_2 - \theta\sigma\sqrt{T}) \end{array} \right\} \qquad (13)$$

with

$K$ = strike price, $S$ = current spot price[5], $T$ = term of the option, $b$ = barrier price, $r$ = risk-free rate, $q$ = asset yield ("deferment rate", in the NNEG context), $\sigma$ = volatility, $\Phi(.)$ is the standard Normal cumulative distribution function, and

$$z_1 = \frac{1}{\sigma\sqrt{T}}\left[\ln\left(\frac{S}{K}\right) + \left(r - q + \frac{\sigma^2}{2}\right)T\right], \qquad z_2 = \frac{1}{\sigma\sqrt{T}}\left[\ln\left(\frac{b^2}{KS}\right) + \left(r - q + \frac{\sigma^2}{2}\right)T\right],$$

$$\theta = 2\frac{(r-q)}{\sigma^2} \quad \text{with } r \neq q.[6]$$

We can make the following observations on Equation 13.

---

[4] For the random walk reflected at zero, see e.g. Equation 2.4 in Gerber and Pafumi (2000); or for the re-transformed density, Equation A.6 in Thomas (2021).

[5] Strictly $S_0$ for consistency with $S_t$ and $S_T$, but I drop the subscript where there is no realistic ambiguity.

[6] For $r = q$ an alternative formula can be derived (Veestraeten 2013, p960). But the formula above suffices for practical purposes, because it remains well-behaved when $r$ is arbitrarily close to $q$, and has only an infinitesimal gap in the solution at $r = q$.

(i) The first line is the Black-Scholes formula[7] for a call, and then there is the $1/\theta$ term (say *Call Adjustment*). The *Call Adjustment* evaluates as positive, but reduces to 0 for $b = 0$, as expected.

(ii) The definition of $S_t$ in Equation 3 excludes the possibility of $S_t$ having touched the barrier in the past, i.e. before the time designated as zero. To be consistent with this set-up, we should always define time 0 as the current time when we do an option valuation, and define the initial notional price as the current price (i.e. $N_0 = S_0$). The density underlying Equation 13 assumes this, i.e. it does not incorporate the possibility of an uprating effect on $S_t$ from past barrier touches, which would require extra terms.

(iii) Some care is needed with the volatility parameter $\sigma$. From Equation 1, we can see that $\sigma$ is the annualised standard deviation of the logarithmic price change of the notional price $N_t$ (i.e. the GBM volatility in the absence of the reflecting barrier). We can estimate $\sigma$ by observing the standard deviation of short-term changes in the observed log price, $\ln(S_t/S_{t-1})$, over a period *in which the barrier has not been touched*, and then applying the usual $\sqrt{t}$-scaling to give an annualised value. But we cannot apply $\sqrt{t}$-scaling to $\sigma$ to give the standard deviation of $\ln(S_T/S_0)$ over a long period $T$, because this will be reduced by the presence of the barrier. The usual $\sqrt{t}$-scaling is predicated on geometric Brownian motion, but that is not what we have in the observed price.[8]

Similarly, for a put, the $\mathbb{Q}_N$-expectation gives a price of

$$P_B = e^{-rT}\mathrm{E}_{\mathbb{Q}_N}\left[K - S_T\right]^+ = e^{-rT}\int_b^K (K - S_T) f(S_T)\, dS_T \qquad (14)$$

which evaluates as

---

[7] Strictly, the Black and Scholes (1973) formula modified for an asset with a continuous yield. But this detail is tedious (every argument in the paper still works if there is no yield), so I shall use "Black-Scholes" throughout.

[8] The variance over interval $T$ of a driftless arithmetic Brownian motion with $\sigma = 1$ is just $T$. With reflection at the starting level, the variance is reduced to $(1 - 2/\pi)T \approx 0.36T$ (Wiersema 2008, p212 gives a proof). Our set-up has the barrier some distance below (rather than equal to) the starting level; this gives a smaller reduction in variance for the log-price.

$$\begin{aligned}
P_B &= Ke^{-rT}\Phi(-z_1 + \sigma\sqrt{T}) - Se^{-qT}\Phi(-z_1) \\
&\quad - be^{-rT}\Phi(-z_3 + \sigma\sqrt{T}) + Se^{-qT}\Phi(-z_3) \\
&\quad + \frac{1}{\theta}\left\{\begin{array}{l} be^{-rT}\Phi(-z_3 + \sigma\sqrt{T}) \\ -Se^{-qT}\left(\frac{b}{S}\right)^{1+\theta}[\Phi(z_4) - \Phi(z_2)] \\ -Ke^{-rT}\left(\frac{b}{K}\right)^{1-\theta}\Phi(z_2 - \theta\sigma\sqrt{T}) \end{array}\right\}
\end{aligned} \quad (15)$$

with

$$z_3 = \frac{1}{\sigma\sqrt{T}}\left[\ln\left(\frac{S}{b}\right) + \left(r - q + \frac{\sigma^2}{2}\right)T\right], \quad z_4 = \frac{1}{\sigma\sqrt{T}}\left[\ln\left(\frac{b}{S}\right) + \left(r - q + \frac{\sigma^2}{2}\right)T\right].$$

For a full derivation, see Hertrich and Zimmermann (2017), supported by Appendices A to C of Hertrich and Zimmermann (2014).

The delta for a call is found by differentiating Equation 13 with respect to the price of the underlying asset:

$$\delta_{C_B} = \frac{\partial C_B}{\partial S} = e^{-qT}\left\{\Phi(z_1) - \left(\frac{b}{S}\right)^{1+\theta}\Phi(z_2)\right\}. \quad (16)$$

Similarly, the delta for a put is:

$$\delta_{P_B} = \frac{\partial P_B}{\partial S} = e^{-qT}\left\{\Phi(z_1) - \Phi(z_3) + \left(\frac{b}{S}\right)^{1+\theta}[\Phi(z_4) - \Phi(z_2)]\right\}. \quad (17)$$

Derivations of Equations 16 and 17 are given in Appendix A.

2.3 Simulation

We now specify some example parameters which will be used where appropriate throughout the rest of the paper: let spot price $S = 1$, barrier $b = 0.5$, strike $K = 1$ (i.e. at-the-money), risk-free rate $r = 0.015$, asset yield $q = 0.01$, volatility $\sigma = 13\%$, and term $T = 25$ years These parameters are intended to be representative of the NNEG context as in Thomas (2021).

Simulation is useful for two distinct purposes: to verify that the formulas are correct $\mathbb{Q}_N$-expectations, and to show that delta-hedging based on these formulas exactly replicates option payoffs.

For the first purpose, we simulate the asset price using Equations 2 and 3, but with the drift $\mu$ replaced by $r$, a fixed large number of time steps, and then vary the number of simulations. We measure pricing error as mean discounted option payoff less the analytical option formula. This is illustrated for a put in Figure 3, a log-log plot of confidence intervals for the pricing error against the number of simulations $N_s$. The slope of - 0.5 shows that the Monte Carlo means converge on the analytical formulas in line with $1/\sqrt{N_S}$.

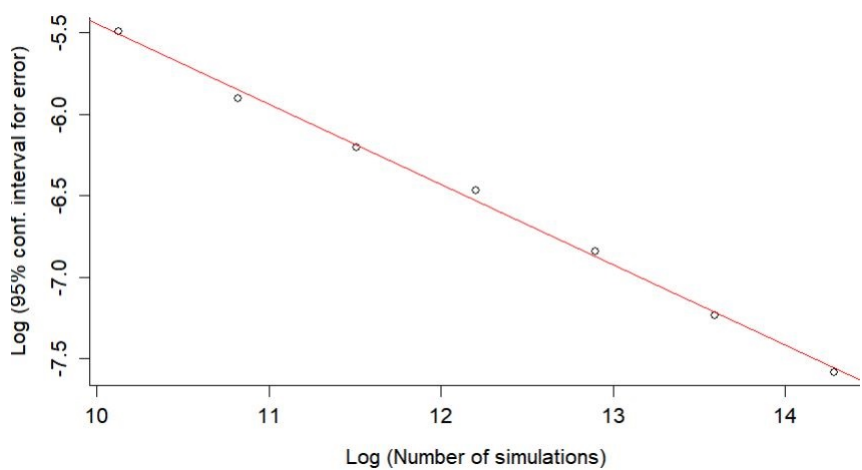

**Figure 3.** Convergence of put pricing errors to zero with increasing number of simulations.

For the second purpose, we again simulate the asset price using Equations 2 and 3, but with a real-world drift $\mu = 0.03$ (say), a fixed number of simulations, and then vary the number of time steps. We find that replication schemes based on initial wealth as per the call and put formulas in Equations 13 and 15, and deltas as in Equations 16 and 17, exactly replicate option payoffs, for any simulated real-world drift of the asset price. This is illustrated for a put in a Figure 4, a log-log plot of standard deviations of replication errors against the number of time steps $N_T$. The slope of - 0.5 shows that the replication errors converge in line with $1/\sqrt{N_T}$.

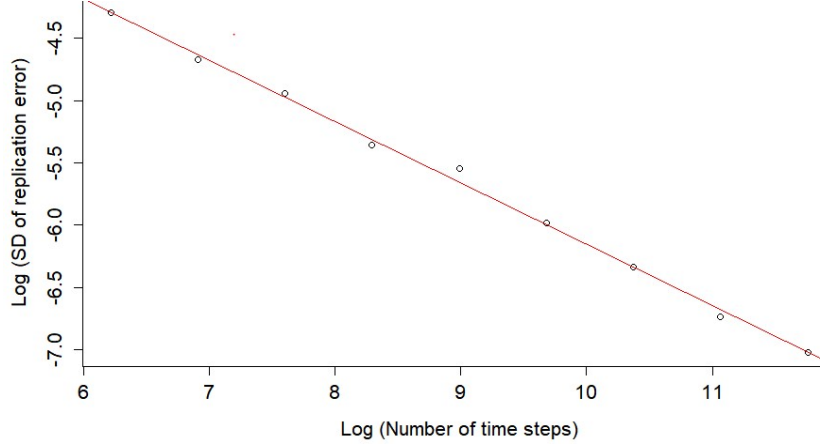

**Figure 4.** Convergence of put replication errors to zero with increasing number of time steps.

2.4 Why delta-hedging exactly replicates option payoffs

Section 2.2 noted that because of the irregularity at the barrier, the observed price $S_t$ is not a martingale under $\mathbb{Q}_N$, and so part of the standard justification for risk-neutral pricing does not apply to options on $S_t$. To understand why option replication by hedging schemes based on $\mathbb{Q}_N$-expectations still works, first recall that the observed price $S_t$ is a function of the notional price $N_t$. Specifically, the definition of $S_t$ in Equation 3 can be written as

$$S_t = N_t \cdot \max\left(1, \max_{0 \le n \le t}\left(\frac{b}{N_n}\right)\right) = N_t . \exp\left\{\max\left[0, \ln b - \ln \min_{0 \le n \le t} N_n\right]\right\} \qquad (18)$$

where the max[.] term on the right is a lookback put (with fixed strike ln $b$) on the log-minimum of $N$. So a put on $S_T$ can be expressed as a compound option on $N_T$, with payoff of

$$\max(K - S_T, 0) = \max\left(K - N_T . \exp\left\{\max\left[0, \ln b - \ln \min_{0 \le n \le T} N_n\right]\right\}, 0\right) \qquad (19)$$

or in words, a put on $S_T$ is a put on: $N_T$ times the exponential of a logarithmic lookback put on the minimum of $N$.[9] As for any derivative of $N_T$, the discounted $\mathbb{Q}_N$-expectation of this payoff is a risk-neutral price with reference to $N_t$; and if $N_t$ were available for hedging, we could replicate its payoff by hedging against $N_t$ using a delta $\partial P_t / \partial N_t$, where $P_t$ is the put price at time $t$. Then note the following points:

---

[9] In the related context of pricing a guarantee on an investment fund, the connection to a lookback option on the minimum was previously noted by Imai and Boyle (2001).

(i) By inspection of Equation 18:

(a) When $N_t$ is not at a new minimum below $b$, then $S_t$ is just a piecewise scaling of $N_t$. Specifically, Equation 18 says $S_t = N_t \tilde{p}_t$, where $\tilde{p}_t$ is the exponentiated payoff at time $t$ of the logarithmic lookback put: $\tilde{p}_t = \exp\left\{\max\left[0, \ln b - \ln \min_{0 \leq n \leq t} N_n\right]\right\}$. The scaling factor, $\tilde{p}_t = \partial S_t / \partial N_t$ for all $S_t \neq b$, is constant throughout each interval when $N_n$ is *not* a new minimum below $b$, corresponding to flat segments of the dashed line in Figure 2.

(b) When $N_t$ hits a new minimum below $b$, substituting $\ln N_t$ for $\ln \min_{0 \leq n \leq t} N_n$ in Equation 18 gives $S_t = N_t (b/N_t) = b$. In other words: whenever $N_t$ hits a new minimum below $b$ (equivalently: $S_t$ touches the barrier), the dependence of $S_t$ on $N_t$ vanishes. So the derivative $\partial S_t / \partial N_t$ must also vanish.

(ii) By the chain rule, $\partial P_t / \partial S_t = \partial P_t / \partial N_t \times \partial N_t / \partial S_t$ (where $P_t$ is the call price evaluated at time $t$). So the option deltas with respect to $S_t$ and $N_t$ are related in the same way as in (i), but with the inverse scaling $1/\tilde{p}_t$, and both must vanish when $S_t$ touches the barrier.

(iii) Hence for the compound option in Equation 19, hedging against $N_t$ using a delta $\partial P_t / \partial N_t$ (which we cannot do) gives the same results as hedging against $S_t$ using a delta $\partial P_t / \partial S_t$ (which we can do).

(iv) By inspection of Equation 17, $\partial P_B / \partial S_t \to 0$ as $S \to b$, as expected from the "vanishing derivative" argument above. The same is true of $\partial C_B / \partial S_t$ in Equation 16.

A further intuition for the vanishing derivatives is as follows. When $N_t$ (a geometric Brownian motion) is near its current lookback minimum, $\tilde{p}_t$, it will almost surely go further below this before time $T$, thus establishing a new minimum. So infinitesimal changes in $\tilde{p}_t$ (and hence $S_t$) at the current time will have no effect on the terminal value of the lookback put, $\tilde{p}_T$ (and hence also, the terminal value of $S_T$). For a rigorous proof see the paper on lookback options by Goldman et al. (1979).

Another way to look at this is via the stochastic process representations of $N_t$ and $S_t$ in Equations 9 and 10. Notice that when $S_t$ is above the barrier, it is just a scaled version of $N_t$, with the same GBM dynamics. When $S_t$ touches the barrier, its dynamics deviate from those of $N_t$, because of the $dL_t$ reflection term; but at this point the option delta goes smoothly to zero, so our replicating portfolio has no exposure to the $dL_t$ term. Stated differently: under the measure $\mathbb{Q}_N$, the discounted option replicating portfolios based on delta-hedging using $S_t$ are martingales; but $S_t$ itself is a sub-martingale, because of the $dL_t$ term.

The above arguments explain why hedging using deltas derived from $\mathbb{Q}_N$-expectations successfully replicates option payoffs, despite the fact that $S_t$ is not a martingale under $\mathbb{Q}_N$. But the difference in dynamics between replicating portfolios (martingales) and asset price (sub-martingale) under $\mathbb{Q}_N$ hints at complications with put-call parity, which we examine next.

## 3. Put-call parity

3.1 Intuition: calls are inefficient, puts efficient

We start with a graphical intuition that foreshadows much of what follows. Briefly, a call is an "inefficient" contract in the sense that its payoff may fail to give credit for the full effect of any interventions when the asset touches the barrier along the asset path to that payoff. A put does not have this feature, and so is "efficient".

To see this, look at Figure 5. Each panel shows one sample path for the notional price $N_t$ (red), and the corresponding observed price $S_t$ (blue). Notice that in the left panel, the payoff of a call, $\max(S_T - K, 0)$, is *less* than the accumulated effect $(S_T - N_T)$ of the interventions (or equivalently, the reflection) along the asset path. In contrast, in the right panel, the payoff of a put, $\max(K - S_T, 0)$, is reduced by the full accumulated effect $(S_T - N_T)$ of the interventions, compared with the much larger payoff if there were no barrier (i.e. if the red line represented the observed price).

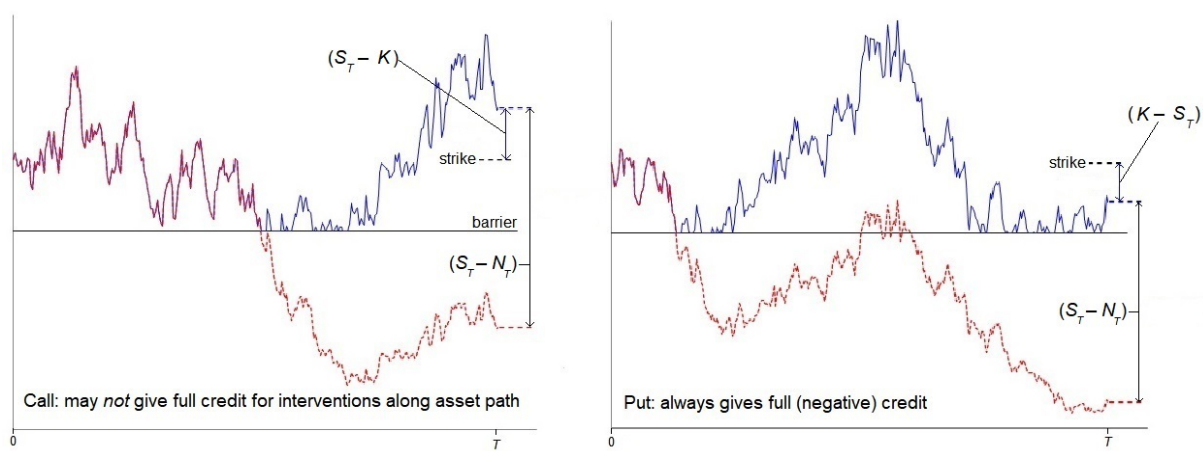

**Figure 5.** Calls are inefficient, puts efficient.

It is possible to draw paths where a call payoff includes the full effect of any interventions, and a put payoff includes none: the red line in each panel just needs to terminate above the strike. But for such paths, the put payoff is zero anyway (both before and after the interventions), so such paths are inconsequential for the valuation of a put. In general, *whenever an option has a positive terminal payoff*:

    (i)    a call payoff *does not* always include the full positive effect (left panel in Figure 5); but

(ii) a put payoff always *does* always include the full negative effect (right panel in Figure 5)

of any interventions along the path to that payoff.

Observations (i) and (ii) suggest that the barrier has, in some sense, a more fundamental effect on a put than on a call.

Also note that:

(a) for a call, the barrier redistributes the density of payoffs compared to that which prevails with no barrier, but without changing its support $[0, \infty]$;

but

(b) for a put, the barrier both redistributes the density of payoffs, and narrows its support from $[0, K]$ to $[0, K - b]$.

Observations (a) and (b) again suggest that the barrier has a more fundamental effect on a put than on a call.

Now recall that by parity, call and put payoffs are mutually redundant: if we have one plus the forward contract, we can always synthesise the payoff of the other. So rather than directly replicating a call, it might be cheaper to synthesise its payoff as a forward plus a put.

3.2 Definition of put-call parity

Instead of the usual general reasoning about payoffs, it is useful for our present purposes to derive put-call parity by explicitly differencing call and put prices:

$$\text{Call price} - \text{Put price} = \text{Forward Contract price} \tag{20}$$

In the Black-Scholes model appropriate for the notional price $N_t$ (a geometric Brownian motion), if the Black-Scholes call and put prices are $C$ and $P$ (assuming the same strike $K$ as the forward contract), we have

$$C - P = \left[ Ne^{-qT}\Phi(z_1) - Ke^{-rT}\Phi(z_1 - \sigma\sqrt{T}) \right] - \left[ Ke^{-rT}\Phi(-z_1 + \sigma\sqrt{T}) - Ne^{-qT}\Phi(-z_1) \right] \tag{21}$$

and then noting that the sums of the Normal functions (e.g. $\Phi(z_1) + \Phi(-z_1)$) each reduce to 1, this gives

$$C - P = Ne^{-qT} - Ke^{-rT} \qquad (22)$$

where the right-hand side corresponds to the no-arbitrage price of a forward contract, as expected. In the Black-Scholes model, Equation 22 holds *at all times* over the term. Stated differently, all of the following are true:

(i) the terminal payoffs of (a) a long call, short put portfolio, and (b) a forward contract, are equal

(ii) the prices of (a) and (b) today are equal

(iii) the replicating portfolios for (a) and (b) are equal at all interim times, and hence have the same dynamics.

I shall refer to points (i) to (iii) as the "triple equivalence" of put-call parity.

For the barrier model, following the same approach of differencing call and put prices (Equations 13 and 15), we obtain

$$C_B - P_B = Se^{-qT}\Phi(z_3) - Ke^{-rT} + be^{-rT}\left(1 - \frac{1}{\theta}\right)\Phi(-z_3 + \sigma\sqrt{T}) + \frac{1}{\theta}\left\{Se^{-qT}\left(\frac{b}{S}\right)^{1+\theta}\Phi(z_4)\right\} \qquad (23)$$

which is *larger* than the standard parity $Se^{-qT} - Ke^{-rT}$, and reduces to that only for $b = 0$.

The right-hand side of Equation 23 can be replicated as the difference of the two replication strategies implied by the left-hand side.[10] As with the constituent options, hedging against the observed price $S_t$ using a $1/\tilde{p}_t$ scaled delta is equivalent to hedging against the notional price $N_t$ using an unscaled delta. The replicating portfolio starts off higher than $Se^{-qT} - Ke^{-rT}$, then evolves as a martingale under $\mathbb{Q}_N$ (the difference of two martingales is itself a martingale) and reaches the payoff of a forward contract $S_T - X$ at maturity. If we denote this "martingale forward contract" as $F_B$, we then have:

$$C_B - P_B = F_B \qquad (24)$$

---

[10] With slight abuse of notation, I shall use $C_B$ and $P_B$ to refer to both replicating strategies and the initial costs of those strategies, as the context requires (and similarly for other replicating strategies, later in the paper).

which satisfies the triple equivalence of put-call parity.

But in the presence of the barrier, there is a simpler and cheaper way of replicating the payoff of a forward contract. Recall the static no-arbitrage argument that underlies the standard forward contract price $Se^{-qT} - Ke^{-rT}$: the payoff of a forward can be replicated by borrowing to buy $e^{-qT}$ units of the asset and continuously reinvesting the income, and this price is enforced because any other price creates a static arbitrage. Then note that in the presence of the barrier, the argument still works in the same way. The static hedge position in the asset means that whenever reflection at the barrier affects the spot price, it will affect the replicating portfolio in the same way. So a static position in the asset, funded by borrowing, still replicates the payoff of a forward contract in the presence of the barrier (and this is easily verified by simulation).

The forward contract defined in this way, say $^{S}F_B = Se^{-qT} - Ke^{-rT}$, has the same terminal payoff as the martingale forward contract $F_B$ in Equations 23 and 24, but a different initial price, and different interim dynamics. Because of the interventions at the barrier, the replicating portfolio for $^{S}F_B$ is a sub-martingale (and hence the $^{S}$-prefix). Note that the cheapness of this replication scheme arises because it fully exploits the interventions at the barrier. In contrast, for the martingale forward contract $F_B$, the replication scheme implied by the right-hand side of Equation 23 has the same property as the call and put replication strategies on the left-hand side: both deltas go to zero as the spot price approaches the barrier, and so the replication scheme gains no benefit from the interventions.

Now recall that the terminal payoff of a call can also be expressed synthetically as a forward contract plus a put, and similarly the terminal payoff of a put can be expressed synthetically as a call less a forward. This suggests two further possible concepts of put-call parity in the presence of the barrier:

replicating the put synthetically:

$$C_B - {}^{S}P_B = {}^{S}F_B \qquad (25)$$

or replicating the call synthetically:

$${}^{S}C_B - P_B = {}^{S}F_B \qquad (26)$$

each of which satisfies the triple equivalence of put-call parity. Other combinations such as $C_B - P_B = {}^{S}F_B$ would *not* satisfy the triple equivalence: the terminal payoffs of the two sides are equal, but the replicating portfolios at earlier times are not. The $^{S}$-prefix in $^{S}C_B$ and $^{S}P_B$ can be thought of as denoting both that the payoffs are replicated synthetically, and that the sub-martingale forward is used in their replication.

The three putative statements of put-call parity in Equations 24, 25 and 26 all reference the same terminal payoff, i.e. they are all the same as regards leg (i) of the triple equivalence of put-call parity. But they are not all the same as regards leg (ii) (current prices) and leg (iii) (interim dynamics). Equations 25 and 26 replicate the same terminal payoff as Equation 24, but using a cheaper replication strategy, which exploits the interventions at the barrier.

3.3 Principle of the lowest price

The fact that replication schemes are not unique does not necessarily mean that all replication schemes are equally sensible. Anyone offering to buy a derivative for more than its *lowest* cost of replication creates a potential arbitrage: sell to the person offering a higher price, replicate the contingent payoff you need to make by setting aside its lowest replication cost, and pocket the difference. And in contexts such as NNEG valuation, it seems reasonable that an insurer should be permitted to value a put at the cost of the cheapest strategy that replicates the payoff, rather than some alternative higher-cost strategy.

This argument suggests the principle that the market price of any derivative should be the cost of the cheapest replication scheme, that is, the scheme which best exploits the interventions at the barrier. We might call this *the principle of the lowest price*, in place of the conventional law of one price. This leaves aside the question of any margin or collateral required by each strategy, to which we shall return below.

3.2      Replication paths

It is now useful to examine some typical simulation paths for the various replicating strategies underlying Equations 24 to 26. Figure 6 shows (clockwise from top left): a single path for $S_t$; the corresponding paths for martingale and sub-martingale forward replications; the direct and synthetic put replications; and the direct and synthetic call replications. I make the following observations (these points are general, despite only a single pair of paths being shown for each comparison).

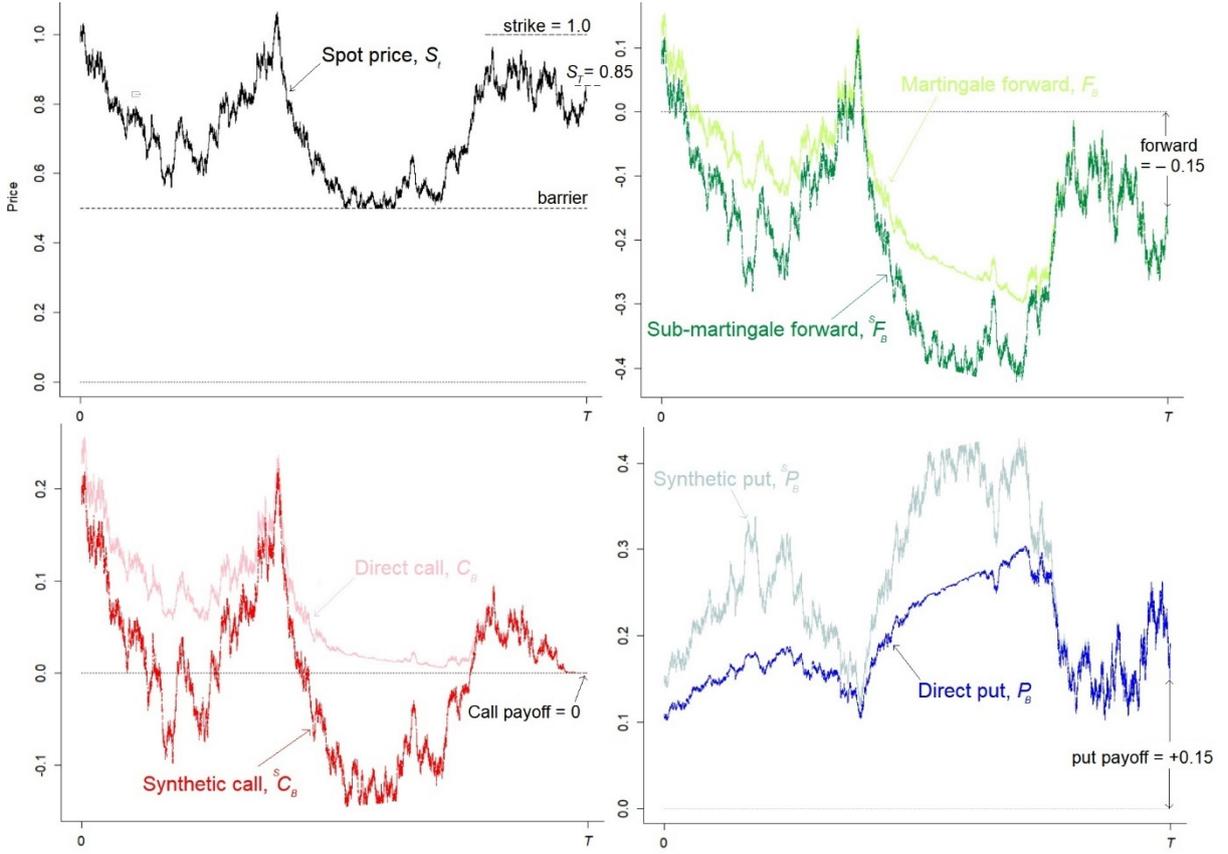

**Figure 6.** Clockwise from top left: paths of spot price, and replicating portfolios for forward, put and call.

(1) In each of the three coloured panels, the difference between the two initial costs, and the subsequent replicating paths, is the same $F_B - {}^S F_B$. In other words, the fundamental point is that there are two ways of replicating the forward contract: $F_B$ which gains no benefit from the interventions, and ${}^S F_B$ which fully exploits the interventions. In each coloured panel, the initial difference is the (discounted) effect of the interventions on the asset price over the whole term. If we rely on the cheaper replicating strategy in each panel, we implicitly anticipate the effect of the free government interventions in our prices.[11] The initial value of the difference is Equation 23 less ${}^S F_B$, that is:

$$F_B - {}^S F_B = be^{-rT}\left(1 - \frac{1}{\theta}\right)\Phi(-z_3 + \sigma\sqrt{T}) - Se^{-qT}\Phi(-z_3) + \frac{1}{\theta}\left\{Se^{-qT}\left(\frac{b}{S}\right)^{1+\theta}\Phi(z_4)\right\} \quad (27)$$

---

[11] We can alternatively extract the value of the interventions over the term, via a suitable delta-trading strategy. This possibility is discussed in Section 6.2.

which is independent of the strike *K*, as expected.

(2)    For a forward contract (green, upper right panel), the sub-martingale replication $^S F_B$ is at all interim times lower than the martingale forward $F_B$. The difference arises from the different replication strategies: $F_B$ misses the interventions, $^S F_B$ captures them. But it might also be rationalised as a reflection of different mark-to-market risks: $F_B$ has a delta of $\delta_{C_B} - \delta_{P_B}$, and $^S F_B$ has a higher delta of $e^{-q(T-t)}$ at interim time *t* (visually, $^S F_B$ has larger drawdowns). Note that $^S F_B$ never exceeds $F_B$; the extra mark-to-market risk is entirely downside risk. On the other hand, $^S F_B$ (a static position) has the advantage that it is much simpler to hedge than $F_B$ (a dynamic position).

(3)    For a put (blue, lower right panel), the direct replication $P_B$ is lower at all interim times than the synthetic replication $^S P_B$. This is intuitive: $P_B$ has no exposure to the interventions at the barrier (because the delta goes to zero); but $^S P_B$ has a short position of $-e^{-q(T-t)}$ (plus reinvestment of income) at the barrier, and so suffers losses from the interventions. If someone offers to buy the put at the higher price $^S P_B$, an arbitrageur can write the put at that price, replicate the contingent payoff he needs to make by setting aside an amount $P_B$, and pocket the difference. Since direct replication is always cheaper, this suggests that a put should be replicated directly.

(4)    Conversely, for a call (red, lower left panel), the synthetic replication $^S C_B$ is lower at all interim times than the direct replication $C_B$. This is again intuitive: $^S C_B$ has a long position of $e^{-q(T-t)}$ when the spot price touches the barrier at interim time *t*, and so fully benefits from all the interventions; but $C_B$ misses out on all the interventions. Since synthetic replication is always cheaper, this suggests that a call should be replicated synthetically (but see the caveat at (6) below).

(5)    Now put the three previous points together, and apply the "principle of the lowest price" articulated in Section 3.2 to each contract. This suggests that we can discard Equations 24 and 25, leaving Equation 26 as the preferred form of put-call parity in the presence of the barrier. That is, a call should always be replicated synthetically, and a put directly; and a forward contract should be replicated by a sub-martingale strategy (i.e. using the same formula as in the absence of the barrier).

(6) However, for a call (but not for a put) there is an important caveat: the red lower left panel in Figure 6 shows that the minimum-cost strategy $^S C_B$ has negative value at interim times during the term. In contrast, the direct replication strategy $C_B$ remains positive at all interim times.[12] So a putative arbitrageur, who sells the call to a person offering the higher price $C_B$ and replicates the contingent payoff he needs to make for $^S C_B$, may need to make margin payments to cover these interim losses. The negative interim value of the strategy does have a lower limit enforced by the barrier, but its magnitude can substantially exceed the initial saving $C_B - {^S C_B}$ plus interest to the interim time.[13] In the presence of credit or margin constraints, this is a possible reason why there may be few or no sellers of the call at the price $^S C_B$, or indeed at any price below $C_B$.

(7) It does not seem possible to give a general resolution of this point: the price of a call is ambiguous, depending on margin requirements in relation to interim losses of the strategy $^S C_B$. Suppose the margin requirements are expressed as a fraction $m$ ($0 \leq m \leq 1$) of the negative interim value of the strategy. We can then resolve the ambiguity for one extreme case: if there are no margin requirements ($m = 0$), it is always cheaper to replicate the call synthetically. On this assumption of "no margin requirements", put-call parity takes the form given in Equation 26:

$$^S C_B - P_B = {^S F_B} \qquad (28)$$

*Synthetic call price – Direct put price = Sub-martingale forward price.*

But more realistically, for any other value of $m$, the best way to replicate a call depends on the availability of credit, and the configuration of $S$, $K$ and $b$ (and hence the maximum possible interim loss on the cheaper strategy $^S C_B$). It might be thought that $m = 1$ (i.e. replication portfolios required to have non-negative value at all interim times) dictates that a call should be replicated directly. But this is not so, because for some configurations of $S$, $K$ and $b$ (particularly $S \rightarrow b$ or $K \rightarrow b$), it may be cheaper to follow the strategy $^S C_B$ and also "pre-fund" the maximum possible margin payment. To see this, first consider a strike equal to the barrier ($K = b$). For this parameterisation, there are no possible interim losses on either strategy $^S C_B$ or $C_B$, so the strategy $^S C_B$ is unambiguously cheaper.[14] Now consider a strike slightly above the barrier. By continuity with the previous case, the maximum possible interim loss of $^S C_B$ is very small; and so following this strategy (which captures the full effect of the interventions) and "pre-funding" the maximum margin payment will be cheaper than following strategy $C_B$ (which completely misses out on the interventions). On the other hand, if either the strike or spot price is well above the barrier (i.e. $K \gg b$ or $S \gg b$), the maximum possible interim loss on the strategy $^S C_B$ is larger; and so for high enough $S$ or $K$, the strategy $C_B$ will be cheaper than $^S C_B$ plus pre-funding of the maximum possible interim loss. Overall, for $m = 1$, the best strategy seems ambiguous: it depends on the configuration of $S$, $K$ and $b$.[15]

---

[12] Note in passing that the issues discussed here never arise in the standard Black-Scholes model, where at all interim times (i) replication portfolios for both calls and puts have positive value and (ii) synthetic replication portfolios have identical value to direct replication portfolios.

[13] Particularly if the barrier at inception is far below the spot price, so that (i) the initial saving $C_B - {^S C_B}$ is small and (ii) the possible interim losses are large.

[14] Stated differently: if $K = b$, a call has the same payoff as a forward, and can be most cheaply replicated as such.

(8) The ambiguity regarding the call as discussed at (6) and (7) does not affect the put. The direct-replication strategy $P_B$ has the lowest initial cost, with any synthetic strategy having higher cost; and $P_B$ never has negative interim value; so $P_B$ always the best replication strategy for a put. This affirms the relevance of $P_B$ for NNEG valuation, notwithstanding the ambiguity about the price of a call.

The next two sections illustrate the principles outlined above, first for the more straightforward case of puts, and then for calls.

**4. Illustration for a put option**

We now return to the example parameters stated in Section 2.3. Figure 7 illustrates the comparative costs of direct replication $P_B$ and synthetic replication $^S P_B$ for a put option, for a range of barrier levels as a fraction of the initial spot price. We can make the following observations on Figure 7.

For a put, direct replication is always cheaper than synthetic replication. To understand this, note that the cheaper replication strategy is the one that is better positioned at the barrier. Then observe that the deltas of the two replication strategies when the spot price touches the barrier at an interim time $t$ are:

synthetic replication: $\delta(C_B - {}^S F_B) = \delta_{C_B} - e^{-q(T-t)} = -e^{-q(T-t)}$

direct replication: $\delta_{P_B} = 0$.

So the synthetic strategy is always badly positioned at the barrier (short the asset, which is subject to positive interventions). This makes it more expensive.

---

[15] There is a loose analogy with the scenario of a bubble in the current price of the asset. Here, Cox and Hobson (2005) and Heston et al. (2007) emphasise that the call price is ambiguous, because it may depend on margin requirements for the replication strategy with lowest initial cost.

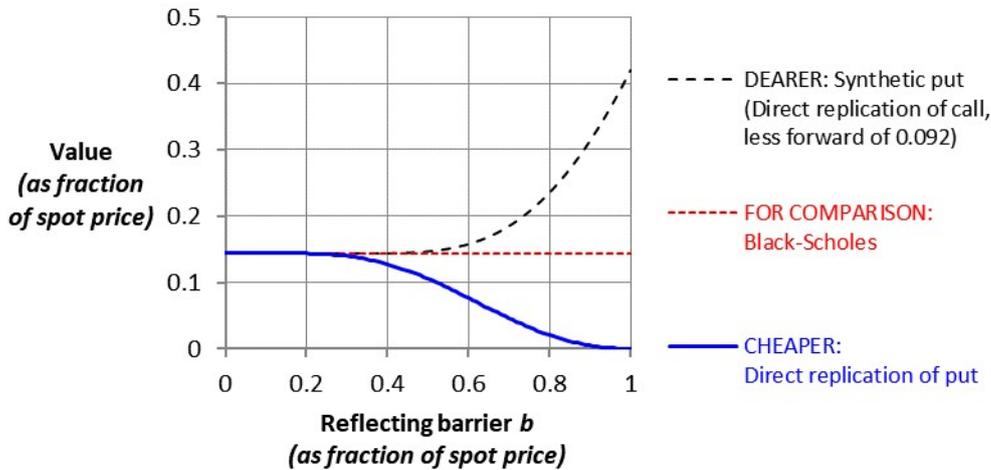

**Figure 7.** Put option: direct replication is cheaper than synthetic replication.

The dashed straight red line representing the Black-Scholes price (i.e. ignoring the barrier) is shown in Figure 7 mainly to illustrate that our preferred valuation, the blue line $P_B$, sensibly converges to the Black-Scholes price for $b = 0$. But perhaps surprisingly, replication based on the Black-Scholes delta will also exactly replicate option payoffs in the presence of a barrier $b > 0$, albeit at unnecessarily high cost. For a discussion of this point, please see Appendix B.

Another way of seeing that a put should be priced by direct replication is to note that a put is a bounded claim with a maximum payoff of $K - b$, and then consider the following extreme cases:

(i)   Barrier equal to strike

For $K = b$, the put payoff becomes zero. In Figure 7, the strike is 1, and the direct-replication price $P_B$ (blue line) goes to zero for $b = 1$, which is intuitively sensible. In contrast, the synthetic-replication price $^S P_B$ (dashed black line) and Black-Scholes price (dashed red line) remain well above zero for $b = 1$.

(ii)   High volatility

For high volatility (e.g. $\sigma > 0.290$ in Figure 7), the synthetic-replication price $^S P_B$ can exceed the maximum possible put payoff $K - b$. But this gives an immediate arbitrage, with no credit requirements: write the put for a cash credit of $^S P_B$, which is larger than the maximum possible payoff you need to make. In contrast, the direct-replication price $P_B$ is well-behaved for high volatility: it increases modestly towards a limit, but remains less than the Black-Scholes price.

## 5. Illustration for a call option

5.1 With no margin requirements: synthetic replication is preferred

Figure 8 illustrates the comparative costs of direct replication $C_B$ and synthetic replication $^S C_B$ for a call option with the same parameters as in Section 4, for a range of barrier levels.

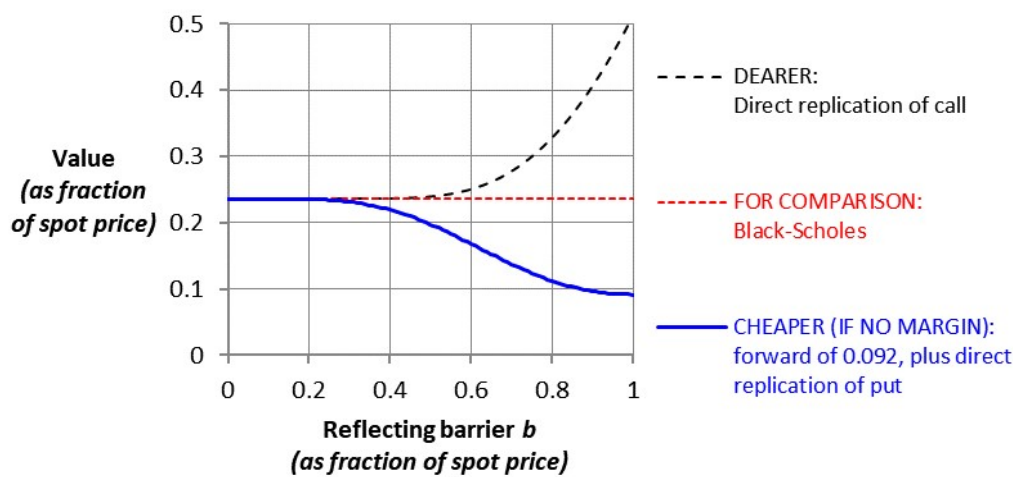

**Figure 8.** Call option: synthetic replication is cheaper than direct replication.

We can make the following observations on Figure 8.

For a call, if we assume $m = 0$ (i.e. no margin payments are required on interim losses of replication), synthetic replication always has a lower initial cost than direct replication. To understand this, note that the deltas of the replication strategies when the spot price touches the barrier at an interim time $t$ are:

synthetic replication: $\delta\left(^S F_B + P_B\right) = e^{-q(T-t)} + \delta_{P_B} = e^{-q(T-t)}$

direct replication: $\delta_{C_B} = 0$

So the synthetic replication strategy is always better positioned at the barrier (long the asset, which is subject to positive interventions).

It may seem counter-intuitive to say that in a world with a *higher* barrier, a call can be replicated at *lower* cost (as a fraction of the initial spot price).[16] To understand this, look at Figure 9. This shows four possible paths for the notional price $N_t$ and the corresponding observed price $S_t$ (in the upper left panel, there are no barrier touches, so the two paths never diverge). In each panel, the effect of the intervention on the terminal price is $S_T - N_T$, and the effect on the call payoff is:

(i) upper left panel: no increase (because no barrier touches, so no interventions)
(ii) upper right panel: increase of $S_T - N_T$;
(iii) lower left panel: increase of *less* than $S_T - N_T$;
(iv) lower right panel: no increase (because terminal observed price below strike).

We can then see that the reduction in call replication cost, compared to a world with no barrier, arises from paths of type (iii). For these paths, the synthetic-replication strategy $^SC_B$ fully captures the effect of the interventions, but the call payoff is increased by *less* than the effect of the interventions. As foreshadowed graphically in Section 3.1, a call is an "inefficient" contract in the presence of the barrier.

We can usefully consider the same extreme cases for the call as we did for the put:

(i) Barrier equal to strike
For $K = b$, the call is economically equivalent to a forward contract. The synthetic-replication price $^SC_B$ then sensibly prices the call as a forward contract, (i.e. $P_B = 0$, so $^SC_B = {^SF_B}$); but the direct-replication price $C_B$ gives an unnecessarily higher price.

(ii) High volatility
For high volatility (e.g. $\sigma > 0.409$ in Figure 8), the direct-replication price $C_B$ exceeds the spot price. This gives an immediate arbitrage, with no credit requirement: write the call and buy the stock for a cash credit $C_B - S$, and then at maturity either deliver the stock and receive the strike (if the call is exercised), or keep the original cash credit (if the call is not exercised). On the other hand, the synthetic-replication price $^SC_B$ sensibly remains less than the spot price for all volatilities.

---

[16] Note that Figure 9 does not say a call is cheaper *in a numeraire of £* in a world with a higher barrier. Value on the y-axis is expressed as a fraction of the initial spot price, which is likely to be higher in a world with a higher barrier.

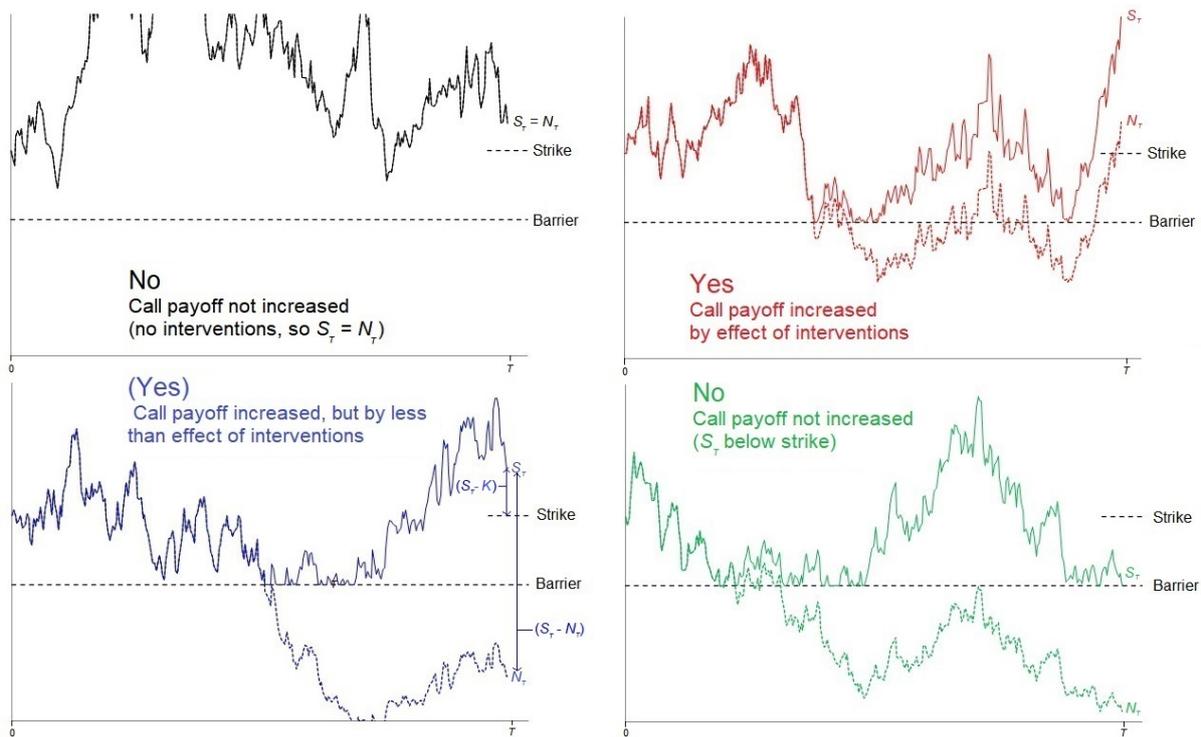

**Figure 9.** Call payoff, $\max(S_T - K, 0)$, may be less than accumulated effect of the interventions.

## 5.2 With margin requirements: call price is ambiguous

The synthetic replication strategy $^{S}C_B$ is subject to the caveat discussed in Section 3, i.e. the interim losses can exceed the initial saving $C_B - {}^{S}C_B$ plus interest to the interim date. This is more likely towards the left of Figure 8, where the initial saving (i.e. the gap between the black and blue curves) is small, and the largest possible interim loss is large. To illustrate, let $V_t\left({}^{S}C_B^t\right)$ be the value of the synthetic replicating portfolio for a call at interim time $t$, and define the "Interim Loss Ratio" (ILR) of the strategy as

$$\text{ILR} = \min_{0 \le t \le T} \left( \frac{V_t\left({}^{S}C_B^t\right)}{\left(C_B - {}^{S}C_B\right)e^{rt}} \right) \qquad (29)$$

where ILR < - 1 indicates that the negative value of the strategy exceeds the rolled-up initial saving at some point during the term. Then on the assumptions at the start of section 2.3 (in particular, $S = K = 1$), for $b = 0.7$, the 25% conditional tail expectation of the ILR is - 0.76 (and none of 10,000 simulations give ILR < - 1). So for $b \ge 0.7$, strategy $^{S}C_B$ might be preferred: the maximum possible interim loss is smaller than the rolled-up difference in initial costs. But for $b = 0.6, 0.5, 0.4$, the corresponding 25% tail expectations of the ILR are - 1.53, - 2.98, - 5.20. For these cases, the strategy $C_B$ might be preferred: it has a slightly higher initial cost, but avoids a substantial chance of interim losses of several times the extra cost.

To summarise this section: ignoring margin requirements, the initial cost of synthetic replication ${}^S C_B$ is always cheaper than that of direct replication $C_B$. But the strategy ${}^S C_B$ may have negative value at interim dates, and these losses can be much larger than the difference in initial costs plus interest to the interim date. If margin payments need to be made to cover interim losses, the price of a call seems ambiguous.

**6. Arbitrage and limits to arbitrage**

The dual replicating strategies for puts, calls and forwards as discussed above point to the possibility for arbitrage strategies in the barrier model. This section examines these possibilities directly, and their practicality in the context of credit constraints.

6.1 Buying when the spot price touches the barrier (not feasible)

Theoretically, if we could buy at the instant when $S_t$ touches the barrier, and sell immediately after reflection, this would be an immediate arbitrage, with no risk of interim losses. However, we have already noted in Section 2.1 that this is not practically feasible: the spot price spends no finite time at the barrier, so no-one can implement the strategy. From a mathematical viewpoint, the existence of this strategy has consequences: it rules out the existence of an equivalent martingale measure for the asset price. But from an economic viewpoint, a purely theoretical arbitrage opportunity seems of no consequence: if no-one can implement the arbitrage, it cannot affect the assumed price process.[17]

Although buying exactly when $S_t$ touches the barrier is not feasible, it may be possible to obtain some sort of approximation. Intuitively, we can do this by increasing a long position as $S_t$ approaches the barrier, and then reducing the position as $S_t$ moves away from the barrier. This type of strategy will be exposed to interim losses, because $S_t$ can fall further after we initiate the position, but the presence of the barrier will place a limit on the losses. All the strategies discussed below amount to some form of this approximation.

---

[17] We could formalise the impracticality by requiring a minimum time interval $h$ between trades. This reflects the practical reality of discrete hedging, and rules out most of the putative purchases exactly at the barrier. It also technically rules out continuous replication strategies; but for small $h$, discrete hedging makes a negligible difference to the accuracy of replication.

6.2 Net-delta strategy (arbitrage with interim losses)

In Figure 6, the pairs of replicating paths for forward, put and call all have the same initial difference, $F_B - {}^S F_B$ (which is independent of the strike $K$), and the differences shrink at the same rate in each panel as the replicating paths converge on the contract payoff. This is because the duality of replicating paths in each panel has the same underlying cause: the forward contract payoff can be replicated either in a cheaper way which exploits the interventions (${}^S F_B$), or in a dearer way which does not ($F_B$). It follows that if we run the cheaper strategy long and the dearer strategy short, we should be able extract the initial difference in prices $F_B - {}^S F_B$ (also equal to $C_B - {}^S C_B$ and ${}^S P_B - P_B$) plus interest over the term. Because the two strikes are equal and opposite, they do not need to be replicated; we can just do the delta trading on each side, funded wholly by borrowing. That is, starting from zero wealth, we borrow to fund a continuously varying "net-delta" position $\Delta$, specified by differentiating minus Equation 27 with respect to the asset price:

$$\Delta = \frac{\partial}{\partial S}\left({}^S F_B - F_B\right) = e^{-qT}\left\{1 - \Phi(z_3) + \left(\frac{b}{S}\right)^{1+\theta}\Phi(z_4)\right\} \tag{30}$$

which is always a long position, and gives the same terminal payoff whether or not the barrier is touched. The terminal payoff is a constant representing the $\mathbb{Q}_N$-expectation of the effect of the interventions on the terminal asset price (i.e. the effect anticipated in our proposed option prices). The strategy reaches its maximum delta and gamma when the asset is at the barrier: it buys at an increasing rate as the asset approaches the barrier and sells at a decreasing rate as it rebounds, and so approximates a purchase exactly at the barrier.

The viability of this strategy is investigated in Appendix C. In summary:

>  (i)  The strategy is difficult to scale if initiated when the spot price is well above the barrier, because the potential interim losses are very large relative to the terminal gain, especially when the strategy is run over shorter terms.
>  (ii) It becomes more scalable if initiated when the spot price is already close to the barrier. This is intuitive: the closer to the barrier a long position is initiated, the less it can lose.
>  (iii) To the extent that the strategy is implemented near the barrier, this may disrupt the assumed price process near the barrier. But this has little effect on the replication of a put, because $\delta_{P_B} \to 0$ as $S \to b$, so that the put replicating portfolio $P_B$ has almost no exposure to the asset price near the barrier.

(iv)     Point (iii) also addresses the more general criticism that even market participants who do not pursue explicit arbitrage strategies may tend to buy as the spot price approaches the barrier, thus modifying the price process in this region. Yes, this may be so; but because the put replication portfolio has almost no exposure to the asset in this region, it makes little difference to the replication of a put.

6.3 Call paradoxes

The synthetic replication strategy for a call captures the full benefit of the interventions at the barrier along all possible paths for the asset, irrespective of the option strike. For a high strike, the $\mathbb{Q}_N$-expectation of the call payoff tends to zero. It follows that the payoff can be funded starting from wealth of *minus* the discounted $\mathbb{Q}_N$-expectation of the interventions. Stated differently: for a high strike, the synthetic-replication strategy $^S C_B$ tends to the net-delta arbitrage strategy in Equation 30, so we can start with wealth of minus the certain gain this will make.

However, the putative "negative price" just suggested assumes that the call writer replicates by the synthetic-replication strategy $^S C_B$. This involves interim losses, which for a high strike can be a large multiple of the negative initial cost (particularly if the initial spot price is well above the barrier, as noted at 6.2(i) above). In contrast, the direct-replication strategy $C_B$ avoids any possibility of interim losses; and as we increase the strike, its initial cost tends to zero from above. As previously noted, the choice between the two strategies (and hence the price of a call) seems ambiguous. It may be reasonable to assume that in practice (a) any offered prices for calls will be strictly positive, but (b) there may be few buyers of calls with high strikes, because the payoff can potentially be replicated more cheaply (albeit with interim losses). In other words, calls with high strikes are likely to be illiquid.

The negative initial cost of replication (ignoring interim losses) for high strikes can also be understood graphically by looking at Figure 9. We have already noted that for paths of the lower left type, the call payoff is less than the quantum of the accumulated interventions, $S_T - N_T$. Now think about what happens as the strike rises. The density of the types of paths shown in the top two panels falls: because the strike is high, very few of the underlying GBM paths for the notional price $N_t$ terminate above the strike. All the call payoff then arises from paths of the advantageous lower left type, where the payoff is more than fully funded by the free interventions.[18]

---

[18] Once again (cf. footnote 15), there is a loose analogy with the scenario of a bubble in the underlying asset. In a bubble, the paradox is that as we increase the strike, the initial cost of replicating a call tends to a positive limit (cf. zero for Black-Scholes); and the standard put-call parity fails (Heston et al. 2007, p381; Cox and Hobson 2005, p478). In a bubble, the current spot price exceeds the discounted expectation (under the pricing measure) of the asset's price at future date, i.e. $S_0 > e^{(r-q)T} \mathrm{E}_{\mathbb{Q}}[S_T]$; with a lower reflecting barrier, this inequality is reversed, so the sign of the limit for the call price is also reversed.

The "high" strike postulated above is, more precisely, high relative to the forward price of $Se^{(r-q)T}$. It follows that a negative call replication cost can also arise for $q > r$, or $S$ sufficiently close to $b$. These effects can be understood as follows.

For $q > r$, a debt-funded long position initiated sufficiently close to the barrier is an arbitrage: the barrier ensures that maximum possible capital loss at maturity is less than the accumulation of the carry $q - r$. Interim losses can occur, because the spot price can move closer to the barrier before much carry has accumulated, but the barrier limits the size of these losses.

For $S \to b$, in terms of Figure 9, if $S_t$ is already close to the barrier, there will be a high density of future asset paths that receive a significant contribution from interventions at the barrier. Most asset price paths that produce positive call payoffs will then be of the advantageous lower left type.

There is a possible economic interpretation of the negative call price. For a sufficiently strong combination of $q > r$, and $S \to b$, a long position in the asset funded wholly by debt can be entered at zero cost, and gives a certain gain (provided that the margin requirement can be financed). A call option may then be a less attractive proposition than the zero-cost long position (especially for higher strikes, which make a positive call payoff less likely).

Any potential exploitation of the paradoxes highlighted in this section implies disruption of the assumed asset price process near the barrier. But as we have already noted in Section 6.2(iv), the put replicating portfolio has almost no exposure to the asset in this region, so the exploitation should make little difference to the replication of a put.

6.4 Bounds on option prices

The modified form of put-call parity suggested in Equation 28:

$$^{S}C_B - P_B = {}^{S}F_B \qquad (31)$$

*Synthetic call price – Direct put price = Sub-martingale forward price*

invalidates the standard lower bound: $\text{Put} \geq \max(Ke^{-rT} - Se^{-qT}, 0)$. This bound is often described as "model-free", but it depends on the standard put-call parity, and is not implied by the modified parity above. To see the invalidity of the standard lower bound, consider a put with a strike equal to the barrier, $K = b$. This put can never have any payoff, and so is sensibly valued at zero, as given by our formula $P_B$; but if $S$ is close to the barrier with $K = b$ and $r < q$, the standard lower bound will be well above zero.

For a call, the standard lower bound is: $\text{Call} \geq \max\left(Se^{-qT} - Ke^{-rT}, 0\right)$. The first limit remains valid in the presence of the barrier: ${}^S C_B \geq Se^{-qT} - Ke^{-rT}$, i.e. a call has optionality, so it is worth at least as much as a forward with the same strike. The second limit of zero is reasonable for offered prices of calls, albeit these will probably be illiquid, as discussed in section 6.3 above.

The inapplicability of the standard lower bound for a put also implies a modification to the standard upper bound for the valuation of an equity release mortgage, as in the "Principle II" promulgated by the Prudential Regulation Authority (2020) in the United Kingdom. This is covered in Appendix D.

## 7. Discussion

7.1 Mathematical versus economic perspectives on arbitrage

When considering the consequences of the various arbitrage possibilities in highlighted in Section 6, it is useful to distinguish between mathematical and economic perspectives.

From a mathematical perspective, the feasibility of any form of arbitrage, even if incapable of practical implementation, negates the "No Free Lunch With Vanishing Risk" condition in the fundamental theorem of asset pricing (Delbaen and Schachermayer, 1994). It follows by the theorem that no equivalent martingale measure for the observed asset price $S_t$ exists; this negates the usual justification for using the computational technique of risk-neutral pricing. But in the barrier model, there is a different justification: we can apply risk-neutral arguments at the level of the notional price $N_t$, (not the observed price $S_t$), but then nevertheless hedge using $S_t$, as explained in Section 2.4.

From an economic perspective, the feasibility of arbitrage suggests that the assumed price process for $S_t$ might not be realistic, insofar as the actions of arbitrageurs may tend to eliminate the features which arbitrage exploits. But the importance of this economic concern depends on the extent to which the arbitrage can be scaled, and whether this eliminates the regularity which it exploits. On this, the following points can be made.

(1)     The arbitrage represents a disequilibrium in the barrier model. But it is a disequilibrium *funded by the government*, and therefore, by assumption, a sustainable one. This sustainability derives from the government's unlimited powers of intervention, as detailed in the next point.

(2)     Whilst the existence of an explicit barrier would imply arbitrage opportunities as identified in Section 6, the barrier is in reality not an explicit feature. Instead, it is a metaphor for the full range of possible policy responses to a large fall in house prices: changes in general monetary and fiscal policies, relaxation of prudential limits on mortgage lending, interest-free and/or non-recourse State loans to house purchasers, regulatory changes which incentivise or compel institutional investors to buy houses, printing money to buy houses, etc. The metaphorical nature of the barrier precludes arbitrage in practice.

(3) The arbitrage opportunities all involve interim losses (except for the impractical one of buying when the spot price is exactly at the barrier). Appendix C shows that the credit requirements are generally substantial relative to the gains, and so the ability to scale the arbitrages will be limited by credit constraints. The credit requirements can be reduced by initiating the strategy only when the spot price is already very close to the barrier (essentially mimicking a purchase exactly at the barrier). But disruption to the price process in this region makes almost no difference to the replication of a put, because the put delta near the barrier is very close to zero.

(4) A number of authors have put forward models where credit constraints, margin requirements and other institutional features lead to the persistence of some arbitrage opportunities in equilibrium, e.g. Shleifer & Vishny (1997), Loewenstein & Willard (2000), Basak and Croitoru (2000), Liu and Longstaff (2004). Others have documented empirical evidence of persistent arbitrages in various markets e.g. Gemmill and Thomas (1997), Lamont and Thaler (2003), Mitchell et al. (2007), Mitchell and Pulvino (2012).

(5) Experience with currency barriers suggests that when the spot price approaches a lower barrier, there is often increased selling (not buying, as the logic of arbitrage would suggest), presumably predicated on speculation that the authorities may not continue to support the barrier. For example, Hertrich and Zimmerman (2017) find that currency option prices shortly before the Swiss National Bank abandoned its lower barrier for the Euro / Swiss Franc rate in January 2015 were consistent with the put formula assuming a "true" barrier significantly below the official policy barrier. To the extent that such speculative selling occurs, it may offset any buying predicated on the logic of arbitrage.

7.2 Alternatives to the barrier model

The complications identified in this paper make the barrier model seem more conditional on practical limits to arbitrage than was apparent in previous papers (Veestraeten 2008, 2013, Hertrich 2015, Hertrich and Zimmermann 2017, Thomas 2021). However, I suggest that its usefulness in applications such as NNEG valuation should be judged by comparison with available alternatives, not an abstract standard of perfection that no extant model attains. If one believes that the barrier assumption, or something like it, is a closer approximation of reality for house prices over the long term than an unrestricted geometric Brownian motion (for reasons, see Thomas 2021, sections 2 and 3), there do not seem to be any wholly satisfactory approaches. Three possible alternatives are as follows.

(i) Standard Black-Scholes, as prescribed by the Prudential Regulatory Authority (2020) for NNEG valuation for certain regulatory purposes. This simply ignores the reality that interventions are likely after large falls in prices, and produces what one well-known writer of long-dated puts has called "wildly inappropriate values when applied to long-dated options" (Buffett, 2011). In the context of NNEG, Black-Scholes seems to me a short-term model stretched well beyond its sensible range of application, like trying to use an egg-timer as a calendar.

(ii) The "real-world" method typically used by UK insurers for NNEG valuation (Hosty et al, 2008). This makes an ad hoc adjustment to the Black-Scholes formula, replacing the forward house price $Se^{(r-q)T}$ with a projected house price $Se^{gT}$, where $g > r - q$. This gives much lower put values than Black-Scholes. If calls are also valued by this method, then to be consistent with parity, the forward price must be $Se^{gT}$ (this can be seen by reworking Equation 21 with $r - q$ replaced by $g$, and is also discussed in Taleb, 2015). The real-world method therefore implicitly assumes the persistence of a large static arbitrage (i.e. parity-implied forward price $Se^{gT}$ less no-arbitrage forward price $Se^{(r-q)T}$). This might be considered a more disconcerting assumption than the persistence of some limited dynamic arbitrage in the barrier model.

To illustrate, in the representative example detailed in Section 2.3 (in particular $S = K = 1$, $b = 0.5$), the value at inception of the arbitrage in the barrier model is 0.0416; this can be extracted only by a dynamic strategy that may incur substantial interim losses. But in the real-world model with a typical house price growth $g = 2.5\%$ and all other parameters the same, the value at inception of the arbitrage is 0.5052 (i.e. 12x larger); this can be extracted by a static strategy with no interim losses.

(iii) The benchmark approach originated by Eckhard Platen (e.g. Platen, 2002; Buhlmann and Platen, 2003; Platen and Heath, 2010; Fergusson and Platen, 2022). This involves a change of numeraire to the "benchmark portfolio" which maximises expected log-utility of terminal wealth (in other literature, the "growth optimal portfolio"). The price of an option is then defined as its real-world conditional expectation in the benchmark numeraire; for long terms, the price is significantly less than Black-Scholes. Effectively, the benchmark approach anticipates the asset's risk premium in general, rather than the prospect of intervention after a large fall in prices. There is no equivalent martingale measure, so some arbitrage is possible, but only with interim losses. There are some apparent paradoxes, but different ones from the barrier model: the price of a put approaches zero when the benchmark portfolio is low, and a modified form of put-call parity applies.

Notably, all of the barrier, real-world and benchmark models allow some form of arbitrage. This perhaps reflects an implicit view that the persistence of some arbitrage opportunities may be a reasonable feature of a long-term asset model, if the ability to scale them in practice is curtailed by credit limits, transaction costs, and other institutional constraints.

## 8. Conclusions

Putting sections 2 to 7 together, we can state the following conclusions.

(1) The observed price process $S_t$ (a reflected geometric Brownian motion) in the barrier model is not arbitrage-free. Arbitrage with interim losses is possible (or with no losses if one could buy exactly at the barrier, but this can never be implemented in practice).

(2) The arbitrage possibilities represent a disequilibrium in the barrier model. But it is a disequilibrium *funded by the government*, and therefore (by assumption) a sustainable one, based on the government's unlimited powers of intervention.

(3) The possibility of arbitrage, whether or not it is implemented, implies (by the Fundamental Theorem of Asset Pricing) that there is no change of measure that will make the observed price $S_t$ a martingale. So the usual justification for risk-neutral option valuation does not apply to options on $S_t$.

(4) Nevertheless, options on $S_t$ can be expressed as compound options on the underlying arbitrage-free notional price $N_t$ (a geometric Brownian motion), to which change of measure and other standard risk-neutral arguments can be applied. Although $N_t$ is not available for hedging, the required hedging schemes for the compound option have deltas which go to zero whenever $N_t$ attains a new minimum below the barrier (equivalently: whenever $S_t$ touches the barrier). Then because, everywhere above the barrier, $S_t$ is just a piecewise-scaled version of $N_t$, hedging with $S_t$ and a piecewise-scaled delta gives the same results as hedging with $N_t$.

(5) The price of a put is clear: direct replication has a lower initial cost than synthetic replication, and the replication portfolio always has positive value, so no extra funding is required for interim losses. This affirms the relevance of the cheaper direct-replication price $P_B$ (rather than the dearer synthetic-replication price $^S P_B$) for NNEG valuation.

(6) The price of a call is ambiguous: synthetic replication has a lower initial cost than direct replication, but the replication portfolio may give interim losses. So the preferred replication strategy, and hence price, of a call may depend on what margin payments need to be made on these losses.

(7) It follows from points (4) to (6) that put-call parity is also in general ambiguous. However, if no margin payments are required on interim losses, we can think of parity as ${}^{S}C_B - P_B = {}^{S}F_B$, that is *Synthetic call price – Direct put price = Sub-martingale forward price.*

(8) Potential arbitrage strategies in the barrier model are all predicated on borrowing to increase a long position as $S_t$ approaches the barrier, and then reducing the position as $S_t$ moves away from the barrier. The practicality of scaling this strategy depends on credit limits and margin requirements. To the extent that arbitrage is implemented (and funded by the government), the assumed price process of reflected geometric Brownian motion may be disrupted, particularly near the barrier.

(9) The delta of the direct replication $P_B$ for a put tends to zero as the asset price approaches the barrier. This makes the strategy more robust to the possible disruption of the assumed price process near the barrier (as mentioned in the previous point).

**Computer code**

Code in the R programming language demonstrating Monte Carlo evaluation of the options and the associated replication schemes is available at https://github.com/guythomas7.

**Acknowledgements**

I wish to thank (without implicating) Markus Hertrich, Alan Reed, and Pradip Tapadar for comments on earlier drafts, and Dean Buckner and Kevin Dowd for useful criticisms of earlier papers.

**Appendix A: Derivation of option deltas**

Please see supplementary material for this appendix.

**Appendix B: Black-Scholes replication in the presence of the barrier**

As well as the direct and synthetic replication strategies discussed in Sections 3 to 5, there is another way of exactly replicating the payoff of a put (or call) on the observed price $S_t$, albeit at unnecessarily high expense. This is to start with wealth equal to the Black-Scholes price and then delta-trade using the Black-Scholes delta (say $\delta_{BS}$) based on the observed price $S_t$. That this approach works in the presence of the barrier may initially seem surprising, but it can be verified by simulation. One way of understanding it is to consider Equation 10, which is reproduced below:

$$dS_t = (\mu - q)S_t\, dt + \sigma S_t\, dW_t + dL_t \qquad (40)$$

Note that whenever $dL_t = 0$, this has the usual Black-Scholes dynamics. Therefore in the initial period before $S_t$ first touches the barrier, the Black-Scholes put replicating portfolio has the same dynamics as in the ordinary Black-Scholes world, and so earns the risk-free rate under $\mathbb{Q}_N$. When $S_t$ touches the barrier, the replicating portfolio has exposure of $\delta_{BS}$ to the asset, and so receives an appropriately scaled infinitesimal increment (or decrement for a put, because the delta is negative). Once $S_t$ departs from the barrier, it resumes its Black-Scholes dynamics. It therefore again earns the risk-free rate under $\mathbb{Q}_N$, until the next time $S_t$ touches the barrier, when another appropriately scaled increment (or decrement for a put) is applied.

Overall, the sequence just described means that the Black-Scholes replicating portfolio for a put starts off at time zero with a value above the discounted $\mathbb{Q}_N$-expectation of the payoff of the put option on $S_T$. The dynamics over the full term – asset price sub-martingale, hence Black-Scholes put replicating portfolio super-martingale – then ensure that the Black-Scholes replicating portfolio converges with the option payoff at maturity.

Black-Scholes is "the barrier model with $b = 0$". By extension of the argument just given, replication using a delta calculated from the barrier model with an assumed barrier $b^*$ anywhere between 0 and the true barrier level $b$ will also exactly replicate option payoffs, albeit at a higher cost than necessary. This can be verified by simulation.

The feasibility of replication with a delta based on any $b^*$ for $0 \leq b^* \leq b$ adds insight into why a delta based on the true level $b$ is the cheapest possible replication strategy for a put. When the spot price touches the barrier, the put replicating portfolio calculated using $b^* < b$ has a negative delta; it therefore incurs an unhelpful decrement from the positive intervention at the barrier. But the delta, and therefore the decrement, are of smaller magnitude than Black-Scholes; and so the initial cost of the replicating portfolio is cheaper than for Black-Scholes. Taking this thought to the limit, the minimal replicating portfolio is achieved by using a delta which reaches zero when the spot price touches the barrier, i.e. a delta based on the true level $b$.

**Appendix C: Net-delta arbitrage and credit limits**

This appendix investigates the practicality of the net-delta arbitrage strategy highlighted in Section 6.2, a continuously varying long position given by:

$$\Delta = e^{-qT}\left\{1-\Phi(z_3)+\left(\frac{b}{S}\right)^{1+\theta}\Phi(z_4)\right\}. \tag{41}$$

I first show that the strategy is costly to scale if initiated when the spot price is well above the barrier. Figure 10 shows results from 1,000 simulations of the strategy over 25 years, initiated when $b/S = 0.5$ (i.e. barrier 50% below spot price, as envisaged in Thomas 2021), and other parameters as in Section 2.3. The left panel shows the distribution of the terminal gain. The right panel shows the lowest cash (i.e. largest borrowing) over the term. The terminal gains lie in a narrow positive band, centred on the difference of the initial forward contract replication costs, $F_B - {}^S F_B$, plus interest over the term (as expected). But the lowest cash over the term has a remarkable bi-modal distribution. It can be seen that strategy involves maximum borrowing of nearly 10x the terminal gain of 0.06 in around 25% of simulations.

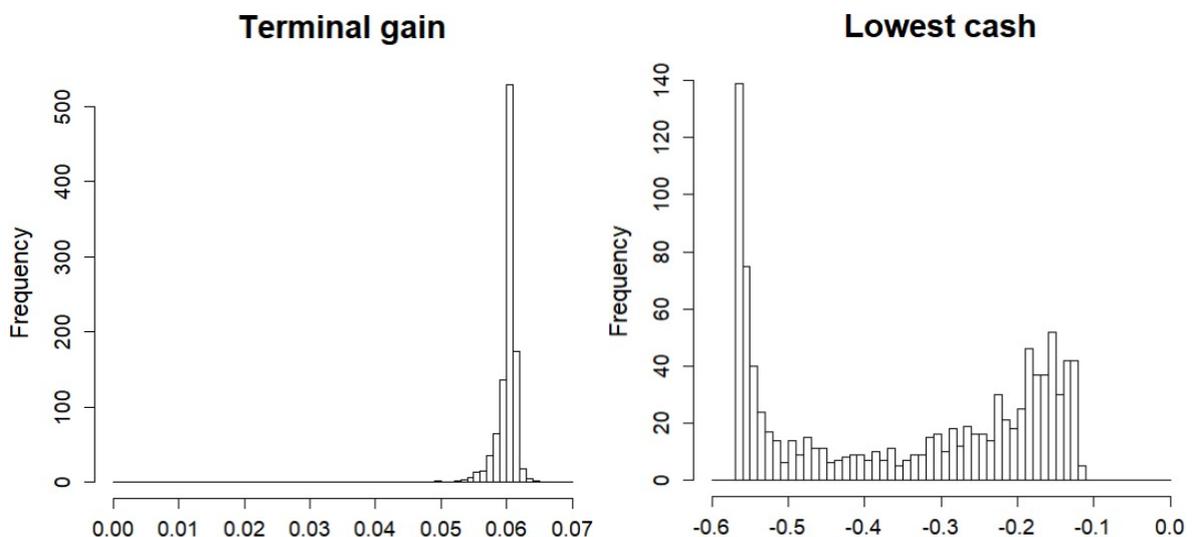

**Figure 10.** 25-year net-delta strategy, initiated when *b*/*S* = 0.5.

To be marketable to third-party investors, the strategy would actually need to be run with a much shorter time horizon (nobody will wait 25 years to see if it works). But when we try this, the drawdowns as a multiple of the terminal gain become much larger, as shown in Figure 11. For a five-year term (perhaps the longest lock-up a third-party investor might countenance), the maximum borrowing is around 600x the terminal gain, and minimum portfolio value (i.e. asset less borrowing) is around minus 100x the terminal gain.

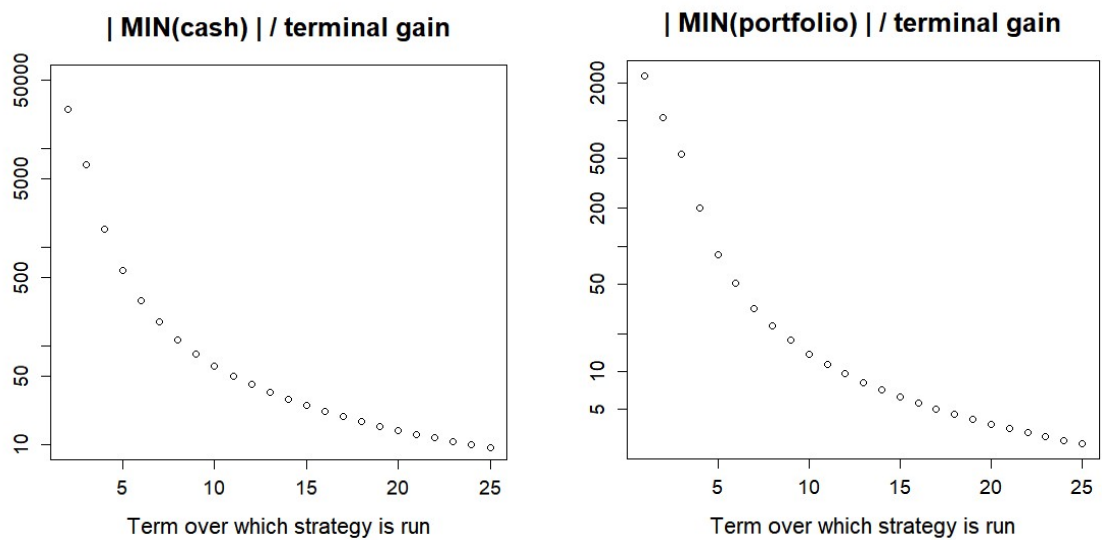

**Figure 11.** Net-delta strategy initiated when *b*/*S* = 0.5, for various terms.

The strategy becomes more attractive if initiated at a time when the spot price is already close to the barrier. Figure 12 shows 1,000 simulations of a one-year version of the strategy, initiated when *b*/*S* = 0.9 to 0.99, with other parameters as in Section 2.3. The maximum borrowing and portfolio value drawdowns as a multiple of terminal gain is now smaller, but still substantial: cash drawdown over 30x for *b*/*S* = 0.9, and 10x for *b*/*S* = 0.99.

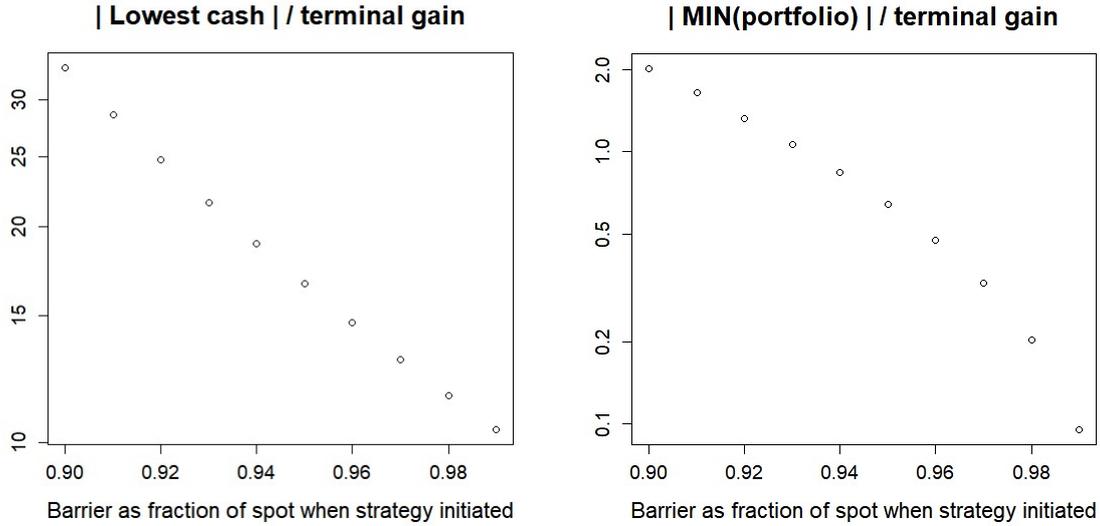

**Figure 12.** One-year net-delta strategy, initiated when $b/S$ = 0.9 to 0.99.

The capital requirement of the strategy reduces to zero only in the limit, where it is initiated when the spot price is exactly at the barrier. But this is unlikely to be possible in any real market, and certainly not for housing.

There remains a more general point: even market participants who do not pursue explicit arbitrage strategies may nevertheless be more inclined to buy when the spot price is close to the barrier, thus potentially modifying the assumed price process in this region. One way of representing this is to draw the log increments for the asset price in a zone near the barrier from a skew-normal distribution, which can be constructed as a mixture of two normal distributions:

$$X_t = \sqrt{1-\alpha^2}\, W_{1,t} + \alpha \left| W_{2,t} \right|, \qquad -1 \leq \alpha \leq 1 \tag{42}$$

where $W_{1,t}$ and $W_{2,t}$ are independent Brownian motions, and $\alpha > 0$ gives a positive skew. Figure 13 illustrates this distribution for two values of the skew parameter, $\alpha = 0.3$ and $\alpha = 0.9$.

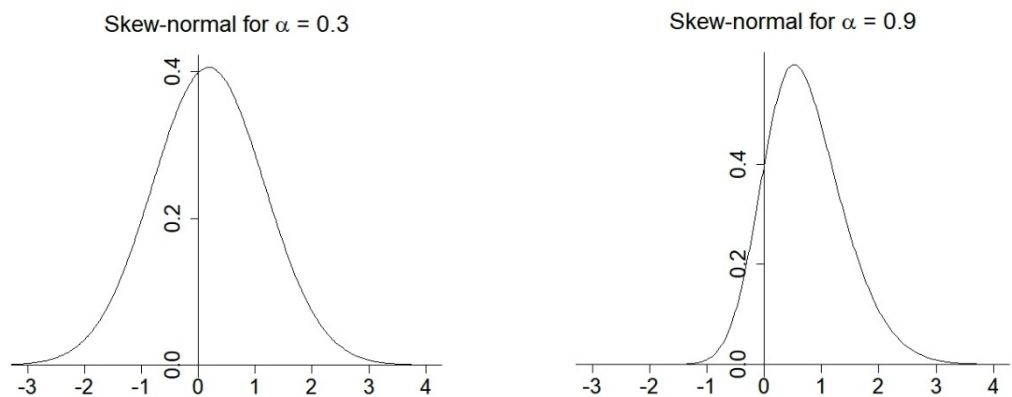

**Figure 13.** Skew-normal distribution.

To represent an increasing preponderance of buying with increasing proximity to the barrier, we can draw increments with a skew parameter $\alpha = 0$ for $S_{t-1} > 1.09b$, $\alpha = 0.1$ for $1.08b < S_{t-1} \leq 1.09b$ and then increase $\alpha$ by a factor of 1.3161 for every 1%-of-$b$ slice below this, reaching a maximum skew of $\alpha = 0.9$ for $S_{t-1} \leq 1.01b$. When we simulate this, the mean Monte Carlo put payoff is reduced (as expected). For parameters as in Section 2.3, (in particular $b = 0.5$, $K = 1$), the reduction is about 6%; but along each path, the lower payoff is almost exactly tracked by our put replication scheme. If we anticipated the skew, we could possibly replicate the payoff for a slightly lower initial cost; to that extent, our formula gives a prudent valuation for a put.

**Appendix D: Modified upper bound for equity release mortgage valuations**

For certain regulatory purposes, the Prudential Regulation Authority (PRA) has promulgated four principles (labelled I to IV) for valuation of an equity release mortgage (ERM) and the NNEG embedded within it. These principles have a common-sense appeal beyond their immediate regulatory context. However, under the assumption of a lower reflecting barrier, part of Principle II may be inappropriate, at least in the form usually stated.

Principle II states[19]:

> *"(II) The economic value of ERM cash flows cannot be greater than either the value of an equivalent loan without an NNEG or the present value of deferred possession of the property providing collateral.*
> *…[This principle] is derived from the following considerations:*
>
> *(i) Given the choice between an ERM and an equivalent loan without an NNEG, a market participant would choose the latter, since either the guarantee is not exercised, in which*

---
[19] PRA (2020), Paragraph 3.15.

*case the ERM and the loan have the same payoff, or it is, in which case the ERM pays less. (ii) Similarly, a market participant would prefer future possession of the property on exit to an ERM, given that the property will be of greater value than the ERM if the guarantee is not exercised, or the same value if it is."*

The verbal presentation of the principle obscures the underlying rationale of the dual limits as an application of put-call parity, which can be seen as follows. Define:

$K_T$    Rolled-up loan at maturity time $T$ (the strike price of the NNEG)

$H_T$    House price observed at maturity time $T$

Then the potential payoff of an ERM in period $T$ (the $T$-period 'ERM-let') is

$$\min(K_T, H_T) = K_T - \max(K_T - H_T, 0) \tag{43}$$

and the present value of this is

$$\text{Present value of } T\text{-period 'ERM-let'} = e^{-rT} K_T - put \tag{44}$$

Noting that *put* is always positive then gives

$$\text{Present value of } T\text{-period 'ERM-let'} \leq e^{-rT} K_T \tag{45}$$

as the first leg of Principle II.

In the absence of a reflecting barrier, applying the lower bound $put \geq \max(K_T e^{-rT} - Se^{-qT}, 0)$ in Equation 44 then gives

$$\text{Present value of } T\text{-period 'ERM-let'} \leq Se^{-qT} \tag{46}$$

as the second leg of Principle II.

But in the presence of a reflecting barrier, the standard lower bound for a put is not sensible, as explained in Section 6.5. So the step at Equation 46 becomes invalid, and hence the second leg of Principle II (i.e. the prepaid forward price $Se^{-qT}$ as an upper limit for ERM) does not apply.

The second leg of Principle II is more likely to be relevant when the term $T$ is large, so that the putative upper bound for ERM given by $Se^{-qT}$ becomes small. This is unchanged by the barrier, and so still becomes small for long terms. But this does not matter to the ERM writer, because the presence of the barrier drives a wedge between the pricing of forwards (including prepaid forwards) and options, and in particular, puts become cheaper to hedge than in the absence of a barrier. By dynamically hedging the put it has written, the ERM writer can be sure of receiving a maturity payment with present value $e^{-rT}K_T - put$ (where the *put* valuation allows for the barrier); the ERM writer does not need to be concerned with the (lower) prepaid forward price.

A generalised form of Principle II to encompass the barrier case might be stated as:

> "The economic value of ERM cash flows cannot be greater than either the value of an equivalent loan without an NNEG, *or any related limit derived from put-call parity*."